\documentclass[twocolumn,secnumarabic,amssymb, nobibnotes, aps, prd]{revtex4-2}

\usepackage{amsmath}
\usepackage{graphicx}
\usepackage{amsfonts}
\usepackage{amssymb}
\usepackage{fancybox}
\usepackage{fancyhdr}
\usepackage{bm}
\usepackage{color}
\usepackage{titlesec}

\newcommand{\be}{\begin{equation}}
\newcommand{\ee}{\end{equation}}
\newcommand{\beq}{\begin{eqnarray}}
\newcommand{\eeq}{\end{eqnarray}}
\newcommand{\n}{\nonumber}
\newcommand{\bk}{\bigskip \noindent}
\newcommand{\sk}{\smallskip \noindent}
\newcommand{\A}{\mathcal{A}}
\newcommand{\B}{\widetilde{\mathcal{B}}}

\newcommand{\F}{\mathcal{F}}
\newcommand{\M}{(M^{-1})}

\newcommand{\D}{\mathcal{D}}
\newcommand{\Dt}{\widetilde{\mathcal{D}}}
\newcommand{\Hh}{\mathcal{H}}
\newcommand{\wD}{\widetilde{\mathcal{D}}}

\newcommand{\wt}{\widetilde{\nabla}}

\newcommand{\R}{\mathcal{R}}
\newcommand{\h}{\sh}
\newcommand{\is}{\int_\Sigma}
\newcommand{\LL}{\mathcal{L}}
\newcommand{\Ss}{\widetilde{S}}

\newcommand{\Xt}{\dot{X}}
\newcommand{\Xtt}{\ddot{X}}
\newcommand{\sh}{\sqrt{h}}
\newcommand{\Gg}{G^{\text{\tiny dyn}}}
\newcommand{\Ggg}{\mathsf{G}^{\text{\tiny dyn}}}

\begin{document}

\title{Hamilton-Jacobi framework for 
Regge-Teitelboim gravity}

\author{Alejandro Aguilar-Salas$^1$, Cuauhtemoc Campuzano$^2$
and Efra\'\i n Rojas$^2$}%
\email{efrojas@uv.mx} 
\email{alejanaguilar@uv.mx} 
\email{ccampuzano@uv.mx}

\affiliation{
$^1$Facultad de Matem\'aticas, Universidad Veracruzana, 
Campus Sur, Paseo No. 112, Desarrollo Habitacional, 
Nuevo Xalapa, 91097, Xalapa-Enr\'\i quez, Veracruz, M\'exico.
\\
$^2$Facultad de F\'\i sica, Universidad Veracruzana, 
Campus Sur, Paseo No. 112, Desarrollo Habitacional, Nuevo 
Xalapa, 91097, Xalapa-Enr\'\i quez, Veracruz, M\'exico.
}

\date{June 21, 2023
\\
{''To the memory of Prof. Riccardo Capovilla}}%

\begin{abstract}
We discuss the Hamilton-Jacobi formalism for brane 
gravity described by the Regge-Teitelboim model, in 
higher co-dimension. Being originally a second-order 
in derivatives singular theory, we analyzed its constraint 
structure by identifying the complete set of 
Hamilton-Jacobi equations, under the Carath\'eodory's 
equivalent Lagrangians method, which goes hand by 
hand with the study of the integrability for this 
type of gravity. Besides, we calculate the 
characteristic equations including the one that 
satisfy the Hamilton principal function $S$. We find 
the presence of involutive and non-involutive 
constraints so that by properly defining a generalized 
bracket, the non-involutive constraints that originally 
arise in our framework, are removed while the set of 
parameters related to the time evolution and the gauge 
transformations, are identified. A detailed comparison 
with a recent Ostrogradsky-Hamilton approach for 
constrained systems, developed for this brane gravity,
is also made. Some remarks on the gauge symmetries 
behind this theory are commented upon
\end{abstract}

\maketitle


\section{Introduction}

The Regge-Teitelboim (RT) gravity theory generalizes general
relativity in a particular theoretical direction~\cite{RT1975}. 
The theory is based on a simple particle/string-inspired
assumption that our Universe is a 4-dim surface isometrically 
embedded, and evolving, in a higher-dimensional flat ambient 
spacetime. On geometrical grounds, our 4-dim spacetime is 
regarded as the trajectory (worldvolume) of a 3-dim extended 
object, floating in a higher-dimensional flat ambient space. 
Within this assumption one must not overlook the local isometric 
embedding theorem~\cite{Friedman1961}. Certainly, the background 
spacetime should be at most 10-dim provided the embedded surface 
be 4-dim. It is known that this number can be lower as long as 
the worldvolume metric admits some Killing vectors. The RT 
action is identical to the Einstein-Hilbert (EH) action of 
general relativity but, with the crucial difference 
that, in this stringy description, the embedding functions 
$X^\mu(x^a)$ rather than the metric $g_{ab}$ defined of the 
worldvolume, are elevated to the level of field variables. 
The field equations result in the Einstein ones but contracted with 
the extrinsic curvature tensor of the embedding which represents 
a generalization of the Einstein equations in the sense that 
the RT equations are fulfilled for every solution of the 
Einstein equations. Under the names of \textit{geodetic 
brane gravity} or \textit{embedding gravity}, the RT model
has been widely investigated by many authors as an interesting
formulation to describe gravity, cosmological implications 
in extra dimensions as well as a suitable alternative for quantization of gravity~\cite{Deser1976,Tapia1989,Maia1989,Davidson1998,
Davidson2003,Pavsic2002,Paston2007,Rojas2009,Paston2010,
Estabrook2010,Sheykin2023,Biswajit2014,Rojas2022,Stern2022,
Paston2023,Stern2023}. The RT model possesses an explicit 
dependence on second order derivatives of $X^\mu$ in a nontrivial 
way. If the aim is to develop a canonical structure for this 
theory, under an appropriate geometric split of the time and 
space derivatives, a linear dependence on the accelerations of 
the extended object can be distinguished while the remaining 
derivatives appear highly entangled; this leads us to the problem 
of not being able to remove an appropriate divergence term that 
leads us to an adequate canonical structure, that is, obtaining 
a quadratic structure of the momenta, which would facilitates its 
quantization. The theory was devised to advance on the quantization 
of gravity since, based on the structural similarity to string theory, 
under the mentioned conditions, a quantization procedure for a 
field theory is more likely to succeed in a flat space. As for 
the last intention it is mandatory, at classical level, to properly 
provide a canonical structure for a given theory and then proceed 
to its quantization. In this sense, in addition to the Hamiltonian 
scheme, the quantization of many physical systems can be related 
with several frameworks and, in particular, with the Hamilton-Jacobi 
(HJ) one.

In view of this, in dealing with the quantization of extended 
objects described by singular models which has been developed 
along different theoretical directions, it seems tempting to 
base the method of quantization of an extended HJ theory for 
continuous and singular physical systems arising from variational 
principles. Several attempts along this line of reasoning have 
been performed in some 
contributions~\cite{Nambu1980,Hosotani1999,Fairlie2001}.

In this paper we examine the geometrical considerations in 
obtaining a Hamilton-Jacobi framework for the RT model, as a 
second-order derivative singular system. This is done on the 
basis of the equivalent Lagrangians method introduced by 
Carath\'eodory~\cite{cara1967} for regular systems and later 
strengthened, from the physical point of view, for first- and 
second-order derivative singular systems by 
G\"{u}ler~\cite{guler1992a,guler1992b,guler1998} and other authors~\cite{pimentel1996,pimentel1998,Muslih1998,hasan2004}.  
To date, there is not a general HJ scheme for this type of 
gravity except when a Friedmann-Robertson-Walker geometry is 
assumed on the worldvolume~\cite{Rojas2020}. In addition to 
this, we find more reasons why we are interested in studying 
and applying this particular HJ framework for extended objects. 
On the one hand, it is not necessary to consider canonical 
transformations derived from the Hamiltonian formalism that 
represents a shortcut  in obtaining a geometric HJ scheme. 
On the other hand, this approach has a deep geometrical meaning 
that establishes that the necessary and sufficient condition 
for an action to be minimized, in a region of the configuration 
space, is the existence of a family of surfaces everywhere 
orthogonal to a congruence of curves. For a singular system, 
with $R$ constraints, such a family arises from solving a set 
of $R+1$ HJ partial differential equations where the complete 
solution is nothing but a congruence of curves, called 
characteristics, that form a dynamical system with 
several independent variables~\cite{cara1967}. Additionally, this 
framework provides an interesting alternative to analyse the 
constraints content and reduced phase space information of the 
physical systems which does not differentiate between first- and 
second-class constraints~\cite{Rojas2022} such that we do not need 
any gauge fixing terms~\cite{Muslih1998}. 
In this spirit, this point of view of HJ replaces the analysis 
of all canonical constraints arising in phase space with the 
analysis of the integrability of the complete set of 
Hamilton-Jacobi partial differential equations of the theory. 
Finally, in this formalism all independent variables labelled 
as $t^I = (t,t^\mu)$, and named parameters of the theory, 
encode the local symmetries and gauge transformations. These 
facts make the approach interesting to initiate a quantization 
program either by using the Feynman path integral scheme or by 
using a WKB approximation~\cite{guler1997,guler1999,Muslih2002,hasan2014}, 
for extended objects and in particular for this type of gravity 
using this HJ formalism.

To reach our intention we will first expose the equivalent 
Lagrangians method for classical field theories under 
the scope of Carath\'eodory, adapted to the geometric notation
describing extended objects. Regarding this, our HJ analysis 
is not covariant. Indeed, in order to explore the predictive 
power mainly for brane-like universes, our description is 
furnished by an ADM decomposition~\cite{Rojas2000,Rojas2004,Tapia1992}. 
Thence we opt to refer to one specific time choice to build the 
HJ constraint analysis. In this sense, we believe that 
for this type of geometrical models the ADM decomposition 
of the theory is more appropriate to pursue the analysis.
These considerations are not immediately applicable for 
any extended object since any geometrical model describing it 
has special features that require a particular discussion. 
Afterwards, we briefly expose the main mechanical characteristics 
of the RT gravity even in the presence of matter. Finally, our 
analysis is then specialized for the RT model as our main issue 
in this work. After determine the HJ partial differential 
equations of the theory, as well as find the characteristic 
equations of the model, we take advantage of the approach and 
turn to analyse the reduced phase space and the gauge content of 
this gravity. In this spirit, following the guideline developed 
in~\cite{Rojas2022}, the comparison with the Ostrogradsky-Hamilton 
approach for RT gravity will be discussed at different stages of 
our analysis.

This paper is organized as follows. In Sec. \ref{sec2}, we derive 
the Hamilton-Jacobi formalism for an extended object described 
by a singular second-order model and its specialization 
to an affine in accelerations theory. In Sec.~\ref{sec3} a 
description of the geometry of the RT gravity and notation used 
is provided. The ADM decomposition of the geometry for extended 
objects and the specialization to the RT gravity are also provided. 
In Sec. \ref{sec4}, we develop the HJ formalism for the RT 
gravity and obtained the full set of Hamiltonians where the 
involutive and non-involutive ones are identified in order to 
build a generalised bracket. Sec.~\ref{sec5} provides a discussion 
of our results. Technical details regarding the brane geometry 
in higher-dimensions, variational derivatives of the Hamiltonians 
of the theory and the Poisson brackets among the Hamiltonians, which
are recurrently used in the paper, are presented in some Appendices.

\section{Hamilton-Jacobi formalism in curved manifolds}
\label{sec2}

The Carath\'eodory's equivalent Lagrangians method to describe 
classical field systems~\cite{guler1998,pimentel1996,pimentel1998}, 
can be extended to analyse singular field theories defined on 
curved manifolds. With a view to analysing theories for extended 
objects described by second-order Lagrangians, $L$, we shall 
consider as field variables the embedding functions $X^\mu$ 
that describe the trajectories swept out by extended objects 
propagating in a background spacetime. When an ADM geometric 
decomposition has been performed for some theory, to be sketched 
below for extended objects, it is appropriate to consider the 
following equivalence between Lagrangian functions
\be 
L' = L - \frac{d \mathsf{S}}{dt}, 
\ee
where 
\be 
\mathsf{S} = \is d^p u\,\sh\,S,
\ee
with $S=S(X^\mu, \dot{X}^\mu, \partial_A X^\mu, t)$ being
a generating function, $\Xt^\mu = \partial_t X^\mu$, 
$\partial_A X^\mu = \partial X^\mu / \partial u^A$ and 
$L = \int_\Sigma d^p u\,\LL$ where $\Sigma$ denotes the 
extended object at a fixed time $t$ and $h=\det (h_{AB})$ 
where $h_{AB}$ is the spatial metric defined on $\Sigma$. 
Here, $u^A$ denotes the space coordinates of 
$\Sigma$~(see \ref{subsec3} below, for more details) and 
$\mu,\nu = 0,1,2,\ldots, N-1$ and $A,B = 1,2, \ldots, p$. 
Under the same ADM framework, $\LL = \LL (X^\mu, \dot{X}^\mu,
\partial_A X^\mu, \ddot{X}^\mu, \partial_A \dot{X}^\mu,\partial_A 
\partial_B X^\mu)$ where the configuration space, $\mathcal{C}_{2N}$, 
is spanned by $X^\mu$ and $\dot{X}^\mu$, under the 
Ostrogradsky-Hamilton framework~\cite{Ostro1850}, where there
are canonical momenta $p_\mu$ and $P_\mu$ conjugate 
to $X^\mu$ and $\Xt^\mu$, respectively. Beyond doubt, the 
variational principle applied to $L' $ leads to the same 
equations of motion.

According to the Carath\'eodory's method, if $L'$ provides 
an extreme of the system then the function $S$ must satisfy the 
relationships given by
\begin{widetext}
\beq 
\frac{\partial (\sh S)}{\partial t} + \left[ \frac{\partial 
(\sh S)}{\partial X^\mu} - \partial_A \left( \frac{\partial 
(\sh S)}{\partial (\partial_A X^\mu)} \right) \right] 
\dot{X}^\mu + \frac{\partial (\sh S)}{\partial \dot{X}^\mu} 
\ddot{X}^\mu - \LL &=& 0,
\label{eq1}
\\
\frac{\partial (\sh S)}{\partial {X}^\mu} - \partial_A 
\left[ \frac{\partial (\sh S)}{\partial (\partial_A X^\mu)} 
\right]
&=& 
\frac{\partial \LL}{\partial 
\dot{X}^\mu} - \partial_t \left( \frac{\partial 
\LL}{\partial \ddot{X}^\mu}\right) - \partial_A \left[ 
\frac{\partial \LL}{\partial (\partial_A \dot{X}^\mu)} 
\right],
\label{eq2}
\\
\frac{\partial (\sh S)}{\partial \dot{X}^\mu} &=& 
\frac{\partial \LL}{\partial 
\ddot{X}^\mu}.
\label{eq3}
\eeq
\end{widetext}
These expressions represent partial differential equations 
(PDE) for the generating function $S$. For regular Lagrangian 
functions, from~(\ref{eq3}) one is able, in principle, to solve 
for $\ddot{X}^\mu$ in terms of $X^\mu$, $\dot{X}^\mu$, $\partial_A 
\dot{X}^\mu$, $\partial_A \partial_B X^\mu$ and partial derivatives 
of $S$. In this manner, it is straightforward to convert~(\ref{eq1}) 
into a PDE for $S$ thus obtaining a HJ framework. 

Nonetheless, for singular systems the previous analysis becomes 
much more complicated, as in the RT case. Since the RT gravity 
is a theory affine in the accelerations~\cite{Rojas2022}, and 
that it is our main interest, from now on we shall restrict
ourselves to analyse a Lagrangian density of the form $\LL = 
\mathcal{K}_\mu \ddot{X}^\mu - \mathcal{V}$ where $\mathcal{K}_\mu$ 
and $\mathcal{V}$ are functions of $X^\mu{}_A,\dot{X}^\mu, X^\mu{}_{AB}$ 
and $ \dot{X}^\mu{}_A$~\cite{Rojas2016,Rojas2021}. The associated 
Hessian matrix is
\be 
\Hh_{\mu\nu} := \frac{\partial^2 \LL}{\partial \Xtt^\mu 
\partial \Xtt^\nu} = 0.
\label{hessian0}
\ee
This entails the obvious vanishing of the determinant of 
$\Hh_{\mu\nu}$ identically. This feature dictates that the 
rank of the Hessian matrix is zero so that the configuration 
space $\mathcal{C}_{2N}$ is fully spanned by $R = N - 0 = N$ 
variables $\dot{X}^\mu$ related with the kernel of 
$\Hh_{\mu\nu}$. In passing, we would like to mention that this 
characteristic is shared by the so-called Lovelock brane 
gravity~\cite{Rojas2013,Rojas2016b}.

\sk
As discussed in~\cite{pimentel1996,Rojas2021} under these 
particular conditions the velocities must play the role of 
free \textit{parameters} of the theory and will be denoted by 
$t^\mu$. In this spirit,  
following the notation of~\cite{pimentel1996}, it 
is feasible to introduce
\be 
t^\mu := \dot{X}^\mu 
\qquad \text{and} \qquad
\Hh_\mu ^P := - \frac{\partial 
\LL}{\partial \ddot{X}^\mu},
\label{eq4}
\ee
which implies the recognition of a whole set of parameters
\be 
t^I :=  (t,t^\mu), \qquad \qquad I = 0,1,2,\ldots, 
N.
\label{eq4a}
\ee
Notice that $\Hh_\mu ^P$ is a spatial density vector of 
weight one. In this spirit, to further simplify the notation, 
it is convenient to set $\Ss := \sh S$. On account of 
definitions~(\ref{eq4}) and~(\ref{eq4a}) it follows 
that~(\ref{eq3}) has the form
\be 
\frac{\partial \Ss}{\partial t^\mu} + \Hh_\mu^P \left( 
t^I, X^\mu, \partial_A X^\mu \right)= 0.
\label{eq5}
\ee

Alike, by introducing the \textit{Hamilton density function}
\be 
\Hh_0 := \left[ \frac{\partial \Ss}{\partial X^\mu} 
- \partial_A \left( \frac{\partial \Ss}{\partial (\partial_A 
X^\mu)} \right) \right] \dot{X}^\mu + \frac{\partial \Ss}{\partial 
\dot{X}^\mu} \ddot{X}^\mu - \LL,
\label{ham1}
\ee
which does not depend on the accelerations $\dot{t}^\mu$, 
the relationship~(\ref{eq1}) becomes
\be 
\frac{\partial \Ss}{\partial t^0} +  \Hh_0 \left(
t^I, X^\mu, \partial_A X^\mu, \frac{\partial \Ss}{\partial 
t^\mu}, \frac{\partial \Ss}{\partial (\partial_A X^\mu)}
\right) = 0,
\label{eq6}
\ee
which is the common structure of the HJ framework. It is also 
evident that we can write~(\ref{eq5}) and~(\ref{eq6}) as 
\be 
\frac{\partial \Ss}{\partial t^I}
+ \Hh_I \left( t^J, X^\mu, \partial_A X^\mu ,\frac{\partial 
\Ss}{\partial t^J},  \frac{\partial \Ss}{\partial 
X^\mu}, \frac{\partial \Ss}{\partial (\partial_A 
X^\mu)} \right) = 0.
\label{HJa}
\ee 
This unified set of $N+1$ PDE for the generating function $S$ 
clearly exhibit the HJ framework we are looking for. Such 
expressions will be referred to as the \textit{Hamilton-Jacobi 
partial differential equations} (HJPDE) of the theory. 

It is convenient to rewrite the system of HJPDE in terms of the 
momenta $p_\mu$ and $P_\mu$. This fact gives rise to a 
suggestive way to work with the system~(\ref{HJa}). Indeed, 
by defining $P_0 := \partial \Ss/\partial t^0$ and noting
that the right-hand sides of~(\ref{eq2}) and~(\ref{eq3})
correspond to the momenta $p_\mu$ and $P_\mu$, \cite{Rojas2004}, 
\beq 
p_\mu &=& \frac{\partial \Ss}{\partial X^\mu}
- \partial_A \left( \frac{\partial \Ss}{\partial \dot{X}^\mu}
\right) = \frac{\delta \mathsf{S}}{\delta X^\mu},
\label{pmuS}
\\
P_\mu &=& \frac{\partial \Ss}{\partial \dot{X}^\mu}
= \frac{\delta \mathsf{S}}{\delta \dot{X}^\mu},
\label{PmuS}
\eeq
respectively, then the system~(\ref{HJa}) becomes
\be 
\Hh' _I (t^J, X^\mu, P_J, p_\mu) := P_I + \Hh_I 
(t^J, X^\mu, P_J, p_\mu) = 0,
\label{HJb}
\ee
where $\Hh'_I := (\Hh'_0, \Hh_\mu^{'P})$ given by~(\ref{eq6})
and~(\ref{eq5}), respectively, and $P_I := (P_0,P_\mu)$.
This strategy has a double duty. On the one hand, it allows 
us to inspect the integrability of the PDE system~(\ref{HJa}) 
while, on the other hand, it enables us to make contact 
with the analysis related to singular theories from the 
viewpoint of gauge systems.

A number of remarks are in order. In view of~(\ref{HJb}), we
recognize the well-known form of the canonical constraints
in the sense of the Dirac-Bergmann approach for constraint
systems~\cite{Dirac1964,henneaux1992,rothe2010}. The HJPDE 
written in the form~(\ref{HJb}) are referred to as 
\textit{Hamiltonians} in this framework. Likewise, 
this HJ scheme replaces the analysis of the genuine $N$ canonical 
constraints  $\Hh_\mu^{'P} = 0$ with the analysis of the $N+1$ 
HJPDE given by (\ref{HJa}). Regarding this, we note 
from~(\ref{eq3}) that we cannot invert $\ddot{X}^\mu$ in terms 
of $X^\mu, \dot{X}^\mu$ and $P_\mu$ such that the definition 
of $P_\mu$ itself entails the presence of $N$ primary 
constraints. In this HJ parlance, $\Hh_0$ is said to be 
associated with the parameter $t$ while the other 
Hamiltonians densities $\Hh_\mu^P$ are associated with the 
parameters $t^\mu$.

At the initial stage the equations of motion in this approach 
are given in terms of the Hamiltonians as
\be 
\begin{aligned}
d X^\mu &= \frac{\partial \Hh'_I}{\partial p_\mu}dt^I
\,\,\,\,\,\,\,\,\qquad 
dt^I = \frac{\partial \Hh'_J}{\partial P_I}dt^J,
\\
d p_\mu &= - \frac{\partial \Hh'_I}{\partial 
X^\mu} dt^I
\qquad 
dP_I = - \frac{\partial \Hh'_J}{\partial t^I} dt^J.
\end{aligned}
\label{eom3}
\ee
Notice the linear dependence on the differentials $dt^I$, the 
parameters of the theory. It is precisely at this point 
where the term Hamiltonians for $\Hh'_\mu$ is justified since 
they generate flows parametrized by $t^\mu$ in analogy with 
the time evolution generated by $\Hh_0$, parametrized by $t$.  
The equations of motion in the form~(\ref{eom3}) are the 
\textit{characteristic equations} (CE) associated with the 
first-order equations system~(\ref{HJa}). The solutions provided 
by~(\ref{eom3}) leads to a congruence of parametrized paths 
given by $X^\mu = X^\mu (t^I)$, and called \textit{characteristic 
curves}. There is another characteristic equation that relates 
the generating function
\beq 
d\Ss &=& \frac{\partial \Ss}{\partial t^I} dt^I +
\left[ \frac{\partial \Ss}{\partial X^\mu} - \partial_A \left( 
\frac{\partial \Ss}{\partial (\partial_A X^\mu)}
\right) \right]dX^\mu ,
\n
\\
&=& - \Hh_I dt^I + p_\mu dX^\mu,
\label{S}
\eeq
where $p_\mu$ given by~(\ref{pmuS}), and~(\ref{HJa}) have been 
considered.

A function $F(u)$ in the phase space, associated with a density 
function $\mathcal{F} (u) = \mathcal{F}(t^I (u),X^\mu (u), 
P_I (u), p_\mu (u))$, will be expressed as the integral 
over the entire spatial hypersurface $\Sigma$, at the time 
$t$, as $F = \is d^p u'\, \mathcal{F} (u')$. In this sense, 
instead of considering Hamiltonian densities as constraints of 
the system, we could promote them as phase space constraint 
functions. Thus, the densities $\Hh'_I$,~(\ref{HJb}), can be recast 
as follows
\be 
H_\lambda = \is d^p u'\, \lambda^I(u') \Hh'_I (u'),
\label{Hfunctions}
\ee
where we have smeared out~(\ref{HJb}) with test fields 
$\lambda^I$ defined on $\Sigma_t$. This idea, based in the 
constraint analysis performed in GR, helps to carry out the 
constraint analysis by avoiding delta distributions in our 
calculations.

For two arbitrary functions $F= \is d^p u \,\mathcal{F}$ and 
$G=\is d^p u \,\mathcal{G}$, defined in the extended phase space, 
$\Gamma$, and spanned by $(t^I,X^\mu)$ and their conjugate momenta 
$(P_I, p_\mu)$, we introduce the \textit{extended Poisson 
bracket} (PB) as
\beq
\{ F, G\} &=& \is d^p u' \left[ \frac{\delta F}{\delta t^I 
(u')} \frac{\delta G}{\delta P_I (u')} + \frac{\delta F}{\delta 
X^\mu (u')} \frac{\delta G}{\delta p_\mu (u')} 
\right. 
\n
\\
&-& \left. ( F \longleftrightarrow G ) \right].
\label{PB}
\eeq
In terms of this, the evolution of any phase space function 
is determined by
\be
d F = \int_\Sigma d^p u'\,\{ F, \Hh'_I (u') \} dt^I (u').
\label{fdiff-1}
\ee
This geometric structure forms, for singular field theories, 
the so-called \textit{fundamental differential} in this HJ
approach.

It is mandatory to analyse the integrability conditions for the 
set of $\Hh'_I$. Indeed, a complete solution 
of the HJPDE~(\ref{HJb}) is provided by a family of surfaces 
orthogonal to the characteristic curves. The conditions that 
ensure the existence of such a family are
\be 
\{ \Hh'_I, \Hh'_J \} = C^K_{IJ}\,\Hh'_K,
\label{frobenius}
\ee
where $C^K_{IJ}$ stands for the structure coefficients of the 
theory~\cite{pimentel2008,pimentel2014}. The conditions~(\ref{frobenius}) 
denote the \textit{Frobenius integrability conditions} from 
the view point of the HJ theory. Thus, under the evolution 
law~(\ref{fdiff-1}), the integrability of the Hamiltonians 
is achieved when the condition  
\be 
d \Hh'_I = \int_\Sigma d^p u\, \{ \Hh'_I, \Hh'_J 
\} dt^J = 0,
\label{fdiff-2}
\ee
holds. This implies that the Hamiltonians must close as 
an involutive algebra which gives rise to calling them 
\textit{involutive}. Whenever there is a bracket among the 
Hamiltonians that cannot be written in the form~(\ref{frobenius}), 
the resulting expression must be considered as a new Hamiltonian 
and has to be incorporated into the original set~(\ref{HJb}) 
and a new complete system can be obtained. The process 
should be repeated until all Hamiltonians, say $\Hh'_{\dot{I}}$, 
satisfy~(\ref{frobenius}) where index $\dot{I}$ will enumerate 
the complete set of parameters $t^{\dot{I}}$. When all the 
Hamiltonians have been found the complete system of HJPDE is 
said to be \textit{complete} and the corresponding system of 
characteristic equations~(\ref{eom3}) is integrable. In this 
process the fundamental differential becomes modified, giving 
rise to a \textit{generalized bracket} (GB) which absorbs the 
non-involutive Hamiltonians~\cite{pimentel2008}.
Taking into account this geometrical structure the resulting 
involutive Hamiltonians generate dynamical evolution, through 
all the arbitrary parameters, that can be understood as a set of 
infinitesimal canonical transformations in the phase space. 
Part of such Hamiltonians serve to identify the gauge 
transformations generator as $G := \Hh'_{\bar{\mu}} \delta 
t^{\bar{\mu}}$,~\cite{pimentel2014}. 
In this work we will elevate the phase space functions 
$H_\lambda$,~(\ref{Hfunctions}), rather than the phase space 
densities $\Hh'_I$, to the level of the geometric structures 
to be handled in order to explore the integrability conditions 
for the RT gravity. 

\section{Regge-Teitelboim gravity}
\label{sec3}

The action functional for RT gravity is the integral over 
the trajectory of $m$, of a $p$-dim brane $\Sigma$, that 
depends on the Ricci scalar $\R$ of $m$ obtained from the
metric $g_{ab}$,~\cite{RT1975},
\be
\A_{\text{\tiny RT}}[X^\mu]= \frac{\alpha}{2}\int_m 
d^{p+1}x\,\sqrt{-g}\, \R ,
\label{RTaction}
\ee
where $\alpha$ is a constant with appropriate dimensions,  
(see Appendix~\ref{app1} for details about of the geometrical 
constructions behind the description of extended objects). 
Possible matter fields living on the brane may be included.
Since we are only interested in the geometrical facts 
we are not coupling brane matter fields and we set $\alpha
=1$. The field variables are the embedding functions $X^\mu 
(x^a)$, where $x^a$ are local coordinates of $m$. Here, 
$\mu = 0,1, \ldots, N-1$ and $a=0,1,\ldots, p$.
The action~(\ref{RTaction}) resembles Einstein-Hilbert action 
defined on $m$, but it is crucial to observe that the field 
variables are the embedding functions and not the components 
of the induced metric.

The most important derivatives of the embedding fields 
$X^\mu$ are encoded in the induced metric and the extrinsic 
curvature 
\be 
g_{ab} = X_a \cdot X_b 
\qquad \text{and} \qquad
K_{ab}^i = - n^i \cdot \nabla_a X_b,
\label{forms}
\ee
respectively, where $\{ e^\mu{}_a := X^\mu{}_a, n^{\mu\,i}\}$ 
with $i=1,2,\ldots, N - p -1$, is the orthonormal basis defined 
on $m$ and $\nabla_a$ is the covariant derivative compatible 
with $g_{ab}$. In fact, $e^\mu{}_a$ denotes the tangent vectors
to $m$. Hereafter, a central dot will denote contraction 
with the Minkowski metric. We assume that the worldvolume is 
without boundary, for simplicity. The symmetries of the action 
are worldvolume reparametrizations, the Poincar\'e symmetry of 
the background Minkowski spacetime and, since we are considering 
higher-codimension, $p+1 < N$, invariance under rotations of the 
normal vectors adapted to the $m$.

Varying~(\ref{RTaction}) under infinitesimal changes
$X^\mu (x^a) \to X^\mu (x^a) + \delta X^\mu (x^a)$, we obtain 
the equations of motion (eom) as a set of conservation laws
\be 
\label{eom1}
\partial_a \left( \sqrt{-g} \,
G^{ab} \partial_b X^\mu \right) = 0,
\ee
or, in a more compact fashion, as a set of $N-(p+1)$  
relationships 
\be 
\label{eom2}
G^{ab} K_{ab}^i = 0,
\ee
where the Gauss-Weingarten equations have been 
used,~\cite{capovilla1995}. Here, $G_{ab}$ denotes the 
worldvolume Einstein tensor. Notice that these equations 
represents a generalization of the Einstein equations 
since the RT equations are fulfilled for every solution 
of the Einstein equations. In fact, expressions~(\ref{eom2}) 
account for the true number of eom. Observe that the eom are of 
second-order in derivatives of $X^\mu$. Of course, we can 
add a matter contribution through a Lagrangian $L_{\text{\tiny 
matt}}$ of fields living on the brane but, such a term does 
not affect the previous geometric forms of the eom. In such 
a case, the eom are given by $(G^{ab} - T^{ab}_{\text{\tiny 
matt}}) K_{ab}^i = 0$ where $T_{ab}^{\text{\tiny matt}} 
= (-2/\sqrt{-g})\partial (\sqrt{-g} L_{\text{\tiny 
matt}})/\partial g^{ab}$.

For brane models described by the extrinsic geometry, the 
conserved stress tensor, $\mathcal{P}_\mu{}^a$,~\cite{Capo2000} 
helps us to determine the mechanical properties of a specific 
model. In our case,
\be 
\mathcal{P}_\mu{}^a = - \sqrt{-g}\,G^{ab}\,e_\mu{}_b,
\label{pamu}
\ee
which is purely tangential to the worldvolume; in fact,
this feature characterize theories leading to second-order
equations of motion. This conserved stress tensor
helps to write~(\ref{eom2}) as a set of projected 
conservation laws 
\be 
(\nabla_a \mathcal{P}^a) \cdot n^i = 0. 
\ee
In passing, the structure~(\ref{pamu}) contracted with the timelike 
unit normal vector $\eta^a$ to $\Sigma$ at a fixed value of $t$, 
allows us to construct the linear momentum density of $\Sigma$ 
as follows
\be 
\pi_\mu := N^{-1} \eta_a \mathcal{P}_\mu{}^a,
\label{pimu}
\ee 
where $N$ represents the lapse function in the ADM decomposition 
for the RT model, as we will see shortly.

\subsection{ADM decomposition of the RT gravity}
\label{subsec3}

It has been described in some contributions the ADM 
decomposition for extended objects as a necessary step 
to describe an appropriate geometrical Hamiltonian 
analysis~\cite{Rojas2000,Rojas2004,Tapia1992}. 
Within such a framework, we start with a slicing of the 
worldvolume $m$ into constant time surfaces $\Sigma_t$, 
each provided with a coordinate system $u^A$ and an 
induced metric $h_{AB}$.

It turns out to be very convenient to describe $\Sigma_t$
by using two related embedding formulations, i.e., as embedded
in $\mathcal{M}$ or as embedded in $m$. A detailed analysis 
is provided in~\cite{Rojas2022}. In Table~\ref{table:geometry} 
we list the main geometric structures  useful in the
ADM formulation for extended objects
\begin{widetext}
\begin{center}
\begin{table}[htbp]
\begin{ruledtabular}
\begin{tabular}{lll}
Embedding & $y^\mu = X^\mu (u^A,t)$ & $x^a = X^a (u^A)$
\\
\hline 
Basis & $\{ \epsilon^\mu{}_A, \eta^\mu, n^\mu{}_i\}$  & 
$\{ \epsilon^a{}_A, \eta^a \}$ 
\\ 
$\Sigma_t$ metric & $h_{AB} =
\eta_{\mu\nu} \epsilon^\mu{}_A \epsilon^\nu{}_B$ & 
$h_{AB} = g_{ab} \epsilon^a{}_A \epsilon^b{}_B$ 
\\
Gauss-Weingarten equations & 
$
\begin{aligned}
\D_A \epsilon^\mu{}_B &= k_{AB}\eta^\mu - L_{AB}^i
n^{\mu\,i}
\\
\D_A \eta^\mu &= k_{AB} h^{BC} \epsilon^\mu{}_C
- L_A{}^i n^\mu{}_i
\\
\D_A n^{\mu\,i} &= L_{AB}^i h^{BC} \epsilon^\mu{}_C
- L_A{}^i \eta^\mu + \sigma_A{}^{ij} n^\mu{}_j
\end{aligned}
$
&
$
\begin{aligned} 
\D_A \epsilon^a{}_B &= k_{AB} \eta^a
\\
\D_A \eta^a &= k_{AB}h^{BC} \epsilon^a{}_C
\end{aligned}
$
\\
Extrinsic curvature & 
$
\begin{aligned}
k_{AB} &= - \eta_{\mu\nu} \eta^\mu 
\D_A \epsilon^\nu{}_B
\\
L_{AB}^i &= - \eta_{\mu\nu}
n^{\mu\,i} \D_A \epsilon^\nu{}_B
\end{aligned}
$ & $k_{AB} = - g_{ab} \eta^a \D_A \epsilon^b{}_B$ 
\\ 
$\Sigma_t$ twist potential & 
$
\begin{aligned}
L_A{}^i &= - \eta_{\mu\nu} n^{\mu\,i}\D_A \eta^\nu
\\
\sigma_A{}^{ij} &= - \eta_{\mu\nu} n^{\mu\,i}
\D_A n^{\nu\,j}
\end{aligned}
$ 
&  
\\
Covariant derivatives & $
\begin{aligned}
\D_A h_{BC} &= 0
\\
\wD_A \phi^i &= \D_A \phi^i
- \sigma_A{}^{ij} \,\phi_j
\end{aligned}
$  & 
$\D_A h_{BC} = 0$
\\
Velocity vector & $\dot{X}^\mu = N \eta^\mu + 
N^A\epsilon^\mu{}_A$ & $\dot{X}^a = N\eta^a + N^A 
\epsilon^a{}_A$
\\ 
ADM induced metric &
$
\begin{aligned}
(g_{ab}) &= \begin{pmatrix}
          -N^2 + N^A N_A & N^Ah_{AB}
          \\
          N^B h_{AB}  & h_{AB}  
           \end{pmatrix}
\\
(g^{ab}) &= \frac{1}{N^2} 
            \begin{pmatrix}
           -1 & N^A
           \\
           N^B & N^2 h^{AB} - N^A N^B
            \end{pmatrix}
\end{aligned}
$
&
\\
ADM extrinsic curvature &
$
(K_{ab}^i) = -
\begin{pmatrix}
n^i \cdot \ddot{X} & n^i \cdot \D_A \dot{X}
\\
n^i \cdot \D_A \dot{X} &  n^i \cdot \D_A \D_B X
\end{pmatrix}
$
&
\\
Worldvolume element 
&
$\sqrt{-g} = N \sh$ 
&
$\sqrt{-g} = N \sh$ 
\\
Projector &
$
\begin{aligned}
\Hh^{\mu\nu} &= h^{AB} \epsilon^\mu{}_A \epsilon^\nu{}_B
\\
&=
\eta^{\mu\nu} + \eta^\mu \eta^\nu - n^{\mu i}n^\nu{}_i
\end{aligned}
$
& 
$
\begin{aligned}
\Hh^{ab} &= h^{AB} \epsilon^a{}_A \epsilon^b{}_B
\\
&= g^{ab} + \eta^a \eta^b
\end{aligned}
$
\\
\end{tabular}
\end{ruledtabular}
\caption{Geometry of $\Sigma_t$ whether it is embedded in
$\mathcal{M}$ or in $m$.}
\label{table:geometry}
\end{table}
\end{center}
\end{widetext}
Further, $N = - \eta \cdot \dot{X}$ and $N^A = h^{AB} 
\epsilon_B\cdot \dot{X}$ denote the lapse function 
and the shift vector defined on $\Sigma_t$, respectively, 
as in the ADM formulation for general relativity.
Additionally, $\D_A$ stands for the torsion-less 
covariant derivative compatible with $h_{AB}$.
It is worthwhile to emphasize that both descriptions
are related by composition; indeed, for example, 
$\epsilon^\mu{}_A = e^\mu{}_a \epsilon^a{}_A$ and 
$\eta^\mu = e^\mu{}_a \eta^a$.

With the geometrical structures given in Table~\ref{table:geometry} 
we are equipped to perform the ADM decomposition of the RT
gravity theory. Indeed, the mean extrinsic curvature, $K^i = 
g^{ab} K_{ab}^i$, turns into
\beq 
\label{Ki}
K^i &=& \frac{1}{N^2}\left[ (n^i \cdot \ddot{X})
- 2N^A (n^i \cdot \D_A \dot{X}) 
\right. 
\n
\\
& & \left.\quad - (N^2h^{AB} - N^A N^B)(n^i \cdot 
\D_A \D_B X)\right].
\eeq
Notice the linear dependence on the accelerations 
$\ddot{X}^\mu$. 
Regarding the Ricci scalar $\R$, by using the 
contracted integrability conditions associated to the 
Gauss-Weingarten equations (see Appendix \ref{app1}), 
we can rewrite it either as a first-order in time derivatives
of $X^\mu$ function plus a total divergence
\be 
\R = R + k_{AB}k^{AB} - k^2 + 2 \nabla_a \left( k \eta^a 
- \eta^b \nabla_b \eta^a \right),
\label{R1}
\ee
where $R$ is the Ricci scalar defined on $\Sigma_t$, 
or alternatively, it can be brought into the suggestive  
second-order in time derivatives of $X^\mu$ function
\beq 
\R &=& 2L_i K^i - G^{ABCD} \,\Pi_{\mu\nu}\, 
\D_A \D_B X^\mu \,\D_C \D_D X^\nu 
\n
\\
&& \qquad \qquad \qquad \quad
- 2 h^{AB} \delta_{ij} \wD_A n^i \cdot \wD_B n^j,
\label{R2}
\eeq 
where $L^i = h^{AB} L_{AB}^i = - n^i \cdot \D^A \D_A X$,
and $\wD_A$ stands for the covariant derivative associated
with the connection $\sigma_A{}^{ij}=\epsilon^a{}_A 
\omega_a{}^{ij}$ that takes into account the rotation freedom 
of the normal vector fields,~\cite{Rojas2022} (see 
Appendix~\ref{app1} for details.). Moreover, we have 
introduced a symmetric normal projector $\Pi_{\mu\nu} 
:= n_\mu{}^i n_{\nu\,i}$ satisfying $\Pi^\alpha{}_\mu 
\Pi^\mu{}_\beta = \Pi^\alpha{}_\beta$, and
\be 
G^{ABCD} := h^{AB}h^{CD} - \frac{1}{2} \left( 
h^{AC}h^{BD} + h^{AD} h^{BC} \right),
\label{wwmetric}
\ee
being a Wheeler-DeWitt like metric associated with $h_{AB}$. 
Whatever the choice for $\R$, the linear dependence on the 
accelerations occur in the divergence 
term in~(\ref{R1}) or through $K^i$ in the first term 
in~(\ref{R2}). In order to maintain a Lagrangian density being 
a covariant scalar we shall use~(\ref{R2}) to construct it 
and then to perform a HJ analysis. This choice will 
preserve the original geometric properties of the model. 

The action functional~(\ref{RTaction}) in terms of an ADM 
Lagrangian reads
\be
\A_{\text{\tiny RT}} [X^\mu] = \int_\mathbb{R} dt 
\,L_{\text{\tiny RT}} (\partial_A X^\mu, \dot{X}^\mu,
\partial_A \partial_B X^\mu, \partial_A \dot{X}^\mu, 
\ddot{X}^\mu),
\label{action2}
\ee
where
\be 
L_{\text{\tiny RT}} = \int_\Sigma d^p u \,
\mathcal{L}_{\text{\tiny RT}},
\label{lag2}
\ee
together with the Lagrangian density  
\beq 
\mathcal{L}_{\text{\tiny RT}} &=& N\h \left( 
L_i K^i - \frac{1}{2} G^{ABCD} \Pi_{\mu\nu} 
\D_A \D_B X^\mu \D_C \D_D X^\nu 
\right.
\n
\\
&-& \left. h^{AB} \delta_{ij} \wD_A n^i \cdot 
\wD_B n^j \right).
\label{lag3}
\eeq
For simplicity, we will introduce the following short 
notation
\beq 
\Theta &:=& G^{ABCD} \Pi_{\mu\nu} \D_A \D_B X^\mu 
\D_C \D_D X^\nu, 
\n
\\
&=& L_i L^i - L_{AB}^i L^{AB}_i,
\label{Theta}
\\
\Psi &:=& h^{AB} \delta_{ij} \wD_A n^i \cdot 
\wD_B n^j,
\n
\\ 
&=& L_{AB}^i L^{AB}_i - L_A{}^i L^A{}_i.
\label{Psi}
\eeq
We must emphasize the linear dependence on the 
accelerations of the Lagrangian density (\ref{lag3}) through 
$K^i$ in the first term. From this fact, it is confirmed that
the Hessian matrix $\Hh_{\mu\nu}$, (\ref{hessian0}), 
associated to~(\ref{lag2}) vanishes identically
\be 
\Hh_{\mu\nu} = \frac{\delta^2 L_{\text{\tiny RT}}}{\delta 
\ddot{X}^\mu \delta \ddot{X}^\nu} = 0,
\label{hessian}
\ee
which identifies a theory affine in accelerations~\cite{Rojas2016}. 
As already mentioned, we will have a fully constrained
phase space.  Regarding the second term in~(\ref{lag3}), it 
involves both the Wheeler-DeWitt like superspace metric and 
the normal projector $\Pi_{\mu\nu}$. Regarding the third term, 
this is precisely the expression of a non-linear sigma 
model built from $n^{\mu\,i}$, with $O (N-p-1)$ symmetry 
that reflects the invariance under rotations of the normal 
vectors $n^{\mu\,i} = n^{\mu\,i} (\dot{X}^\nu)$, constrained 
to satisfy $n^i \cdot n^j = \delta^{ij}$. 

Since we are considering the RT model as a second-order 
theory, the Ostrogradsky-Hamilton framework is an 
appropriate alternative when one does not want to 
alter the original geometric information \cite{Ostro1850}. 
In this sense, we have a $4N$-dim phase space spanned 
by two conjugate pairs $\{ X^\mu, p_\mu; \dot{X}^\mu, 
P_\mu \}$ where the momenta, conjugate to $\dot{X}^\mu$ 
and $X^\mu$, in terms of the $\Sigma_t$ basis are
\beq 
P_\mu &=& \frac{\delta L_{\text{\tiny RT}}}{\delta 
\Xtt^\mu}
= \frac{\h}{N}L^i n_{\mu\,i},
\label{P}
\\
p_\mu &=& \frac{\delta L_{\text{\tiny RT}}}{\delta 
\Xt^\mu} 
= \pi_\mu + \partial_A (N^A P_\mu + \h \,h^{AB} 
L_B{}^i n_{\mu\,i}),\quad 
\label{p}
\eeq
respectively, where $\pi_\mu$ is given by~(\ref{pimu}).
One must notice that~(\ref{P}) and~(\ref{p}) are spatial
densities of weight one, a fact originating from the 
presence of the factor $\h$. As discussed in~\cite{Rojas2022},
according to the Legendre transformation $\Hh_0 = p \cdot 
\dot{X} + P \cdot \ddot{X} - \LL$, the associated canonical 
Hamiltonian density reads
\beq
\Hh_0 &=& p \cdot \dot{X} + 2N^A (P \cdot \D_A \dot{X})
\n
\\
&+& (N^2 h^{AB} - N^A N^B ) (P \cdot \D_A \D_B X)
\n
\\
&+& \frac{1}{2} N\sh \, \Theta + N \sh \,\Psi,
\label{ham0}
\eeq
where we considered~(\ref{Theta}) and~(\ref{Psi}). 
It is worth to notice the linear dependence on the 
momenta $p_\mu$ and $P_\mu$. In this sense, the absence 
of a quadratic term $P^2$ in~(\ref{ham0}), might be a signal 
that we are dealing with a non-authentic second-order derivative 
model~\cite{Nesterenko1988,Rojas2004}.

\section{HJ analysis for Regge-Teitelboim gravity}
\label{sec4}

Based in the spacetime foliation performed in GR, by means 
of an ADM decomposition for extended objects~\cite{Rojas2000,Rojas2004,Tapia1992}, the HJ 
analysis we are going to build is not complete covariant. 
According to the expression for the momenta~(\ref{P}) 
and the canonical Hamiltonian~(\ref{ham0}), and taking into account~(\ref{HJb}), we immediately identify the initial set of 
Hamiltonians in the HJ sense
\beq 
\Hh'_0 &=& P_0 + \Hh_0 = 0,
\label{hami0}
\\
\Hh^{'P}_\mu &=& P_\mu - \frac{\sh}{N} L^i 
n_{\mu\,i}
=0.
\label{hami1}
\eeq

As was remarked above, it is mandatory to test the 
integrability for these Hamiltonians, which within
the HJ scheme translates into analysing the integrability 
of the HJPDE. To accomplish this, it is convenient to 
project~(\ref{hami1}) along the basis $\{ \dot{X}^\mu, 
\epsilon^\mu{}_A, n^{\mu\,i} \}$ to obtain the totally 
equivalent set of $N$ Hamiltonians
\beq 
\Hh^{'P}_1 &=& P \cdot \dot{X} 
= 0,
\label{hami1a}
\\
\Hh^{'P}_A &=& P \cdot \partial_A X 
= 0,
\label{hami1b}
\\
\Hh^{'P}_i &=& P \cdot n_i - \frac{\sh}{N} L_i = 0.
\label{hami1c}
\eeq
This equivalence is sustained by using the projector 
$\Hh^{\mu\nu}$ in the bulk onto $\Sigma_t$ in $\Hh^{'P}_\mu 
= \eta_{\mu\nu} \Hh^{'P\,\nu}$,  see Table~\ref{table:geometry}. 
The adopted route is accompanied by the splitting of the 
original parameters $t^\mu$  labelled as follows: $t^\mu 
\longrightarrow (t, t^1, t^A, \omega^i)$. In this sense, 
$t, t^1, t^A$ and $\omega^i$ are the parameters
associated to the Hamiltonian densities $\Hh_0' ,
\Hh_1^{'P}, \Hh_A^{'P}$ and $\Hh_i^{'P}$, respectively.

In view of~(\ref{Hfunctions}), we wish to address the 
integrability analysis through the phase space constraint 
functions
\beq 
H^{'P}_\lambda &:=& \is d^pu\, \lambda \,(P \cdot 
\dot{X}),
\label{cons1}
\\
H^{'P}_{\vec{\lambda}} &:=& \is d^pu\, \lambda^A 
(P \cdot \partial_A X),
\label{consA}
\\
H^{'P}_{\vec{\phi}} &:=& \is d^pu\,\phi^i \left( 
P \cdot n_i - \frac{\sh}{N} L_i \right),
\label{consi} 
\\
H^{'p}_{\Lambda} &:=& \is d^pu \, \Lambda \left( P_0 
+ \Hh_0 \right),
\label{cons0}
\eeq
where $\lambda, \lambda^A$, $\phi^i$ and $\Lambda$ 
are generic test fields defined on $\Sigma_t$. 
Hereafter, as a simplification in the notation, 
the differential $d^p u$ wherever a $\Sigma$
integration is performed will be absorbed. We also 
introduce the phase space function
\beq 
\Gg &:=& H'_{dt} + H^{'P}_{dt^1} + H^{'P}_{d\vec{t}} 
+ H^{'P}_{d\vec{\omega}},
\n
\\
&=& \is  \left( 
\Hh'_0 \,dt + \Hh^{'P}_1 dt^1 + \Hh^{'P}_A dt^A 
+ \Hh^{'P}_i d\omega^i
\right),\,\,\,\,\,\quad
\label{G}
\eeq
which is nothing but the differential generator of 
dynamical evolution in $\Gamma$. Therefore, for any 
phase space function $F$, the dynamical evolution can 
be written as
\be 
dF = \{ F, \Gg \}.
\label{dFG}
\ee
In the following we will repeatedly perform partial 
integrations and we shall proceed to remove boundary 
terms without altering the symplectic structure.

We turn now to test the integrability conditions for 
the primary Hamiltonians~(\ref{hami1a}),~(\ref{hami1b}) 
and~(\ref{hami1c}). 
We begin with
\beq 
d H^{'P}_\lambda &=& \{ H^{'P}_\lambda, H'_{dt} \} 
+ \{ H^{'P}_\lambda , H^{'P}_{dt^1} \}
+ \{ H^{'P}_\lambda , H^{'P}_{d\vec{t}} \}
\n
\\
&+&  \{ H^{'P}_\lambda, H^{'P}_{d\vec{\omega}} \},
\n
\\
&=& - \is \lambda\,\Hh_0\,dt,
\n
\eeq
where we have made use of the corresponding functional 
derivatives and the extended Poisson brackets listed in
Appendices \ref{app2} and~\ref{AppC}. We then have
\be 
\Hh_0 = 0,
\label{hami2a}
\ee
that is, the canonical Hamiltonian vanishes identically.
This result was expected due to the invariance under 
reparametrizations of this field theory. In view of 
this fact, one is tempting to ensure, with support 
on~(\ref{hami0}), that $P_0 = 0$ must be imposed as a 
new Hamiltonian but, this is not the case since this 
feature is mandatory to respect the worldvolume 
reparametrization invariance which translates into 
the vanishing of the partial derivatives with respect 
to $t$ of quantities defined on $\Sigma_t$. Second, 
we have
\beq 
d H^{'P}_{\vec{\lambda}} &=& \{ H^{'P}_{\vec{\lambda}}, 
H'_{dt} \} + \{ H^{'P}_{\vec{\lambda}} , H^{'P}_{dt^1} \}
+ \{ H^{'P}_{\vec{\lambda}} , H^{'P}_{d\vec{t}} \}
\n
\\
&+& \{ H^{'P}_{\vec{\lambda}}, H^{'P}_{d\vec{\omega}} \},
\n
\\
&=& - \is  \lambda^A (p \cdot \partial_A X 
+ P \cdot \partial_A \dot{X} )\,dt.
\n
\eeq
In view of~(\ref{fdiff-1}) and~(\ref{fdiff-2}) we find 
that
\be 
\Hh^{'p}_A :=  p \cdot \partial_A X 
+ P \cdot \partial_A \dot{X} = 0,
\label{hami2b}
\ee
is a new set of $p$ Hamiltonian densities. Third, the 
condition $d H^{'P}_{\vec{\phi}}=0$ in view of~(\ref{fdiff-1}) 
and~(\ref{fdiff-2}), yields
\beq 
d H^{'P}_{\vec{\phi}} &=& \{ H^{'P}_{\vec{\phi}}, 
H'_{dt} \} + \{ H^{'P}_{\vec{\phi}} , 
H^{'P}_{dt^1} \} + \{ H^{'P}_{\vec{\phi}} , 
H^{'P}_{d\vec{t}} \}
\n
\\
&+& \{ H^{'P}_{\vec{\phi}}, H^{'P}_{d\vec{\omega}} \},
\n
\\
&=& - \is \phi^i n_i \cdot \left[ p  
- \partial_A \left( N^A P 
+ \sh h^{AB}L_B{}^i n_j \right) 
\right]dt
\n
\eeq
from which we infer that
\be 
\Hh^{'p}_i := p \cdot n_i - n_i \cdot \partial_A 
\left( N^A P + \sh L^A{}_j \,n^j \right) = 0,
\label{hami2c}
\ee
is also a new set of $N-p-1$ Hamiltonian densities. 
Finally, 
\beq 
d H^{'p}_\Lambda &=&  \{ H^{'p}_\Lambda, H'_{dt} \} 
+ \{ H^{'p}_\Lambda , H^{'P}_{dt^1} \}
+ \{ H^{'p}_\Lambda , H^{'P}_{d\vec{t}} \}
\n
\\
&+& \{ H^{'p}_\Lambda, H^{'P}_{d\vec{\omega}} \},
\n
\\
&=& 0.
\n
\eeq

Now, the integrability conditions of the Hamiltonians~(\ref{hami2a}),~(\ref{hami2b})
and~(\ref{hami2c}) must be tested. To do this, 
according to our strategy, by smearing out~(\ref{hami2b}) 
and~(\ref{hami2c}) with test fields $\Lambda^A$ and 
$\Phi^i$, respectively, we introduce the phase space 
constraint functions
\beq 
H^{'p}_{\vec{\Lambda}} &:=& \is \Lambda^A
\left( p \cdot \partial_A X + P \cdot \partial_A \dot{X} 
\right),
\label{HLvp}
\\
H^{'p}_{\vec{\Phi}} &:=& \is \Phi^i \left[  p \cdot n_i
- n_i \cdot \partial_A \left(
N^A P + \sh L^A{}_j \,n^j  \right) \right].
\,\,\,\,\,\,\,\,\,\,
\label{HPhivp}
\eeq
According to the general theory, it is required to 
add a new set of parameters associated with the new 
Hamiltonians. This entails complementing the generator 
of dynamical evolution~(\ref{G}) as follows
\beq 
\Ggg &:=& H^{'p}_{dt} + H^{'P}_{dt^1} + 
H^{'P}_{d\vec{t}} + H^{'P}_{d\vec{\omega}} + 
H^{'p}_{d\vec{\tau}} + H^{'p}_{d\vec{\chi}},
\label{GG}
\eeq
where the parameters $\tau^A$ and $\chi^i$ are related to the 
Hamiltonian densities $\Hh^{'p}_A$ and $\Hh^{'p}_i$,~(\ref{hami2b}) 
and~(\ref{hami2c}), respectively. Thus, the new 
fundamental differential reads
\be 
dF = \{  F , \Ggg \}.
\label{DfGG}
\ee
From this, by imposing $d H^{'p}_{\vec{\Lambda}} = 0$ we 
obtain
\beq 
d  H^{'p}_{\vec{\Lambda}} &=& 
\{ H^{'p}_{\vec{\Lambda}}, H'_{dt} \} 
+ \{ H^{'p}_{\vec{\Lambda}}, H^{'P}_{dt^1} \} 
+ \{ H^{'p}_{\vec{\Lambda}}, H^{'P}_{d\vec{t}} \}
\n
\\
&+& \{ H^{'p}_{\vec{\Lambda}}, H^{'P}_{d\vec{\omega}} \}
+ \{ H^{'p}_{\vec{\Lambda}}, H^{'p}_{d\vec{\tau}} \}
+ \{ H^{'p}_{\vec{\Lambda}}, H^{'p}_{d\vec{\chi}} \},
\n
\\
&=& 
0.
\label{cond1}
\eeq
In carrying out the calculation, specifically in the 
last Poisson bracket, we find an advantage in using 
the geometric structure
\be 
\label{BAi}
\B^A{}_i := \frac{\h}{N} (N^A L_i + N L^A{}_i),
\ee
in addition to the use of the integrability 
conditions (\ref{int-cond}) as well as the identity 
$[\Dt_A , \Dt_B ] \Phi^i = \Omega_{AB}{}^{ij}\Phi_j$.
The main reason to introduce (\ref{BAi}) is based in 
the boundary term appearing in (\ref{HPhivp}). 
Certainly, because of the definition for $P_\mu$,~(\ref{P}), 
contracted along the normal vectors $n^\mu{}_i$, the 
phase space constraint~(\ref{HPhivp}) can also be 
written in the form,
\be 
H^{'p}_{\vec{\Phi}} = \is \Phi^i \left( p \cdot n_i
- \wD_A  \B^A{}_i \right),
\n
\ee
where we have neglected a surface term. Hence, 
$d  H^{'p}_{\vec{\Lambda}} = 0$ is fulfilled 
identically.

We deduce similarly, by imposing $d H^{'p}_{\vec{\Phi}} 
= 0$, the relationship
\beq 
d H^{'p}_{\vec{\Phi}} &=& \{ H^{'p}_{\vec{\Phi}}, 
H'_{dt} \} + \{ H^{'p}_{\vec{\Phi}}, H^{'P}_{dt^1} \} 
+ \{ H^{'p}_{\vec{\Phi}}, H^{'P}_{d\vec{t}} \}
\n
\\
&+& \{ H^{'p}_{\vec{\Phi}}, H^{'P}_{d\vec{\omega}} \}
+ \{ H^{'p}_{\vec{\Phi}}, H^{'p}_{d\vec{\tau}} \}
+ \{ H^{'p}_{\vec{\Phi}}, H^{'p}_{d\vec{\chi}} \},
\n
\\
&=& \{ H^{'p}_{\vec{\Phi}} , H'_{dt} \}
+ \{ H^{'p}_{\vec{\Phi}}, H^{'P}_{d\vec{\omega}} \}
+ \{ H^{'p}_{\vec{\Phi}}, H^{'p}_{d\vec{\chi}} \},
\n
\eeq
By considering the PB
\beq
\A:= \{ H^{'P}_{\vec{\phi}}, H^{'p}_{\vec{\Phi}} \}
&=& \is  \phi^i \widetilde{\A}_{ij} \Phi^j,
\label{eq61}
\\
\F := \{ H^{'p}_{\vec{\Phi}}, H^{'p}_{\vec{\Phi}'} \}
&=& \is d^p u \, 
\widetilde{\F}^A_{ij} (\Phi^i \wD_A \Phi^{'j} 
\n
\\
&& \qquad \qquad \qquad
- \Phi^{'\,i} \wD_A \Phi^j),
\label{eq62}
\\
&=&  \is  \widetilde{\F}_i (\Phi^j, \wD_A \Phi^j) 
\,\Phi^{'\,i},
\\
\mathcal{J} := \{   H^{'p}_\Lambda , H^{'p}_{\vec{\Phi}}  \}
&=&  \is  \Lambda \,\widetilde{\mathcal{J}} (\Phi^i, \wD_A 
\Phi^i),
\label{eq60} 
\eeq
where $\widetilde{\A}_{ij}$, $\widetilde{\F}^A_{ij}$,
$\widetilde{\F}_i$ and 
$\widetilde{\mathcal{J}}$, are awkward structures of the 
coordinates $u$ and are listed in Appendix~\ref{AppC}, 
we get
\be
d  H^{'p}_{\vec{\Phi}}
= \is 
\left( - \widetilde{\mathcal{J}}\,dt - \phi^i 
\widetilde{\A}_{ij}\,d\omega^j - \widetilde{\F}_i\,d\chi^i 
\right) = 0.
\label{cond2}
\ee
Notice that $\mathcal{J} = \mathcal{J} (\Lambda, \Phi^i)$.
Within this HJ framework, condition~(\ref{cond2}) displays the 
dependence of the parameters $\omega^i$ and $\chi^i$ in terms 
of the time parameter $t$. As was emphasized above, it is 
possible to eliminate such a dependence by introducing an 
appropriate generalized bracket \cite{pimentel2008,pimentel2014}. 
To do so, we first rename some Hamiltonian 
functions as follows
\be 
\n
\begin{aligned}
h'_1 &:= H^{'P}_{\vec{\phi}}
\\
h'_2 &:= H^{'p}_{\vec{\Phi}}
\end{aligned}
\qquad \longrightarrow \qquad h'_{\bar{I}},
\quad \bar{I},\bar{J} = 1,2.
\ee
Then, we turn to build the matrix elements $M_{\bar{I}\bar{J}}:= 
\{ h'_{\bar{I}}, h'_{\bar{J}} \}$ and its inverse matrix 
components, $\M^{\bar{I}\bar{J}}$.
In matrix form they are given by
\be 
(M_{\bar{I}\bar{J}}) = 
\begin{pmatrix}
0 & \A
\\
-\A & \F
\end{pmatrix}
\,\,\, \text{and} \,\,\,\,
(\M^{\bar{I}\bar{J}}) = \frac{1}{\A^2} 
\begin{pmatrix}
\F & -\A
\\
\A & 0
\end{pmatrix},
\ee 
respectively.

We define the generalized bracket (GB) for this 
theory as
\be 
\{ F, G \}^* := \{ F, G \} - \{ F, h'_{\bar{I} }\} 
\M^{\bar{I}\bar{J}} \{ h'_{\bar{J}}, G \}.
\label{GB}
\ee
This structure applied to all the Hamiltonians of the 
theory leads us to a closed algebra. Therefore~(\ref{hami1a}), ~(\ref{hami1b}),~(\ref{hami2a}) and~(\ref{hami2b}) furnish 
us with the complete set of involutive Hamiltonian densities 
under the GB~(\ref{GB}) and determine a subspace in the 
phase space where the theory is integrable.

We have thus determined that the evolution in phase space is 
provided by the fundamental differential
\be 
dF = \{ F, \mathbb{G} \}^*,
\label{dF}
\ee
where 
\be 
\mathbb{G} := H^{'p}_{dt} + H^{'P}_{dt^1} 
+ H^{'P}_{d\vec{t}} + H^{'p}_{d\vec{\tau}}.
\label{generator}
\ee
It is worthwhile to note that the evolution at the reduced
phase space depends on $2(p+1)$ parameters and the theory is 
now complete and integrable. In other words, (\ref{generator})
is the generator of the characteristic flows along the directions
of the independent parameters $t,t^1,t^A$ and $\tau^A$.

\subsection{Characteristic equations}
\label{subsec4}

Taking into account~(\ref{dF}) we turn now to compute the 
characteristic equations. These include the time evolution
behaviour and the relationships related to the gauge 
symmetries of the theory. The canonical 
Hamiltonian,~(\ref{hami2a}), belongs to the set of 
involutive Hamiltonians. Apparently, to analyze the gauge
transformations, we must turn off the performance of $\Hh_0$ 
due to its relationship to the time parameter $t$, but 
this will lead us to a contradiction. Indeed, following the 
discussion in GR, this constraint plays a double duty since it 
generates diffeomorphisms out of the spatial hypersurface. On 
the one hand, from the background spacetime point of view 
this fact represents a gauge transformation whereas for a 
spatial observer this is dynamics so that we can not cleanly 
separate what is gauge from what is 
dynamics~\cite{Wald1984,Ashtekar1988}. In this sense, it 
will be necessary to write~(\ref{generator}) according to 
the specific point of view that we are interested on.
  
In order to reproduce the equations of motion,
it is mandatory to identify the form of the parameters.
In this spirit, $\mathbb{G}$ is considered as defined 
in (\ref{generator}) with the labeling $\mathbb{G} \to 
\mathbb{G}^{\text{\tiny dyn}}$. 
Hence, the first characteristic equation reads
\beq 
dX^\mu &=& \{ X^\mu , \mathbb{G}^{\text{\tiny dyn}} \}^*,
\n
\\
&=&  \{ X^\mu , \mathbb{G}^{\text{\tiny dyn}} \}
- \{ X^\mu, h'_2 \} \M^{21} \{ h'_1, 
\mathbb{G}^{\text{\tiny dyn}} \},
\n
\\
&=& \dot{X}^\mu \,dt + \epsilon^\mu{}_A\,d\tau^A
\eeq
which exhibit the independence between the parameters $t$ 
and $\tau^A$; in this sense, $d\tau^A/dt = 0$. The second 
characteristic equation for $dt^\mu$ should reproduce the 
form of the accelerations of the brane as well as helps 
to identify some of the parameters of the theory. Certainly,
\beq
d\Xt^\mu &=& \{ t^\mu, \mathbb{G}^{\text{\tiny 
dyn}} \}^*
\n
\\
&=& \{ t^\mu, \mathbb{G}^{\text{\tiny 
dyn}} \} - \{ t^\mu, h'_1 \} \M^{12} \{ h'_2, 
\mathbb{G}^{\text{\tiny dyn}} \},
\n
\\
&=& \left[ 2N^A \D_A \Xt^\mu + (N^2 h^{AB} 
- N^A N^B) \D_A \D_B X^\mu \right]dt
\n 
\\
&+& \Xt^\mu dt^1 + \epsilon^\mu{}_A dt^A
+ \D_A \Xt^\mu d\tau^A + \frac{\mathbb{A}}{\A}
\phi^i n^\mu{}_i \,dt,
\n
\eeq
where $\mathbb{A}:= \is \widetilde{\A} (\Phi^i, 
\D_A \Phi^i)$, and thus we are able to construct the 
accelerations of the brane
\beq
\Xtt^\mu &=& 2N^A \D_A \Xt^\mu + (N^2 h^{AB} 
- N^A N^B) \D_A \D_B X^\mu 
\n
\\
&+& \Xt^\mu \frac{dt^1}{dt} + \epsilon^\mu{}_A 
\frac{dt^A}{dt} 
+ \frac{\mathbb{A}}{\A} \phi^i n^\mu{}_i,
\label{id10}
\eeq
where we have considered the fact that $d\tau^A / dt = 0$. 
With support of the identity that relates the acceleration 
of the brane in terms of the $\Sigma_t$ basis~\cite{Rojas2004}
\beq 
\Xtt^\mu &=& ( \dot{N}_A + N \D_A N - N^B \D_A N_B) 
\epsilon^{\mu\,A} + (\dot{N}
\n
\\
&+& N^A \D_A N  + N^A N^B k_{AB})\eta^\mu + (n^i \cdot \Xtt)n^\mu{}_i,
\n
\eeq
appropriately contracting~(\ref{id10}) with the $\Sigma_t$ 
basis we are able to fix the derivatives of the parameters 
as follows
\beq 
\frac{dt^1}{dt} &=& \frac{\dot{N}}{N} - \frac{N^A}{N} \D_A N
- Nk,
\label{id11}
\\
\frac{dt^A}{dt} &=& N \partial_t\left( \frac{N^A}{N}\right)
+ N \D_B \left( Nh^{AB} - \frac{N^A N^B}{N}\right)
\n
\\
&+& N^A \D_B N^B - N N^A k,
\label{id12}
\\
\frac{\mathbb{A}}{\A} \phi^i &=& N^2 K^i.
\label{id13}
\eeq
Inserting~(\ref{id11}),~(\ref{id12}) and~(\ref{id13}) into 
the characteristic equations $dP_\mu$ and $dp_\mu$, after a 
cumbersome but straightforward computation, following the same 
calculation strategy exposed in~\cite{Rojas2022}, we are able to 
reproduce the definition of $p_\mu$,~(\ref{p}), and the equations of 
motion~(\ref{eom1}), (or~(\ref{eom2})), respectively. This is 
actually the dynamical evolution of the RT gravity. In passing, we 
would like to mention that our results concerning the evolution
in phase space, in comparison with those obtained in~\cite{Rojas2022} 
by using a Dirac-Bergmann scheme for constrained
systems, are the same so we are clearly in agreement. 

On the other hand, when $\{ X^\mu, \Xt^\mu \}$ and  $\{ p_\mu, 
P_\mu \}$ are considered as pure gauge degrees of freedom that merely 
represent changes of coordinates, then $\mathbb{G}$ is specialized
to consider arbitrary parameters, i.e., $\mathbb{G} \to \mathbb{G}^{\text{\tiny can}} = H^{'p}_{\delta t} + H^{'P}_{\delta \vec{t}} 
+ H^{'P}_{\delta t^1} + H^{'p}_{\delta \vec{\tau}}$. The arbitrary deformations of phase space variables result
\begin{widetext}
\beq 
\delta_G X^\mu &=& \{ X^\mu, \mathbb{G}^{\text{\tiny 
can}} \}^*,
\n
\\ 
&=& \Xt^\mu \,\delta t + \partial_A X^\mu \,\delta \tau^A,
\label{G1}
\\
\delta_G \Xt^\mu &=&  \{ X^\mu, \mathbb{G}^{\text{\tiny 
can}} \}^*,
\n
\\ 
&=&
\left[ 2N^A \D_A \Xt^\mu + (N^2 h^{AB}  - N^A N^B) \D_A \D_B X^\mu 
\right] \delta t + \partial_A \Xt^\mu \,\delta \tau^A
+ \Xt^\mu \,\delta t^1 
+ \partial_A X^\mu \,\delta t^A
-  \frac{\mathcal{J}(\delta t, \Phi^i)}{\A} 
\phi^i n^\mu{}_i,\,\,\,\,
\label{G2}
\\
\delta_G \,P_\mu &=& \{ P_\mu, \mathbb{G}^{\text{\tiny 
can}} \}^*,
\n
\\ 
&=&
- p_\mu\, \delta t -2 (P \cdot \D_A \Xt)\epsilon_\mu{}^A 
\,\delta t 
+ 2N^A (P \cdot \D_A \D_B X) \epsilon_\mu{}^B\,\delta t
+ 2N (P \cdot \D_A \D^A X) \eta_\mu \delta t
+ \h \left( \frac{1}{2} \Theta + \Psi \right)\eta_\mu 
\delta t
\n
\\
&-& \h \,G^{ABCD} L_{AB}^i k_{CD}\, n_{\mu\,i}\,\delta t
+ \D_A (2N^A P_\mu\,\delta t)
+ 2 \frac{\h}{N} \left[ \wD_A \left( NL^A{}_i\,\delta t\right) 
- N L^{AB}_i k_{AB}\,\delta t \right]n_{\mu\,i}
\n
\\
&-& \frac{\mathcal{J}(\delta t^0, \Phi^i)}{\A} \frac{\h}{N^2}
\phi^i ( L_i \,\eta_\mu +  k \,n_{\mu\,i}),
\label{G3}
\\
\delta_G \,p_\mu &=& \{ p_\mu, \mathbb{G}^{\text{\tiny 
can}} \}^*,
\n
\\ 
&=& \D_A \left[ p_\mu \,\delta \tau^A + P_\mu \,\delta t^A
- \widetilde{T}_\mu{}^A (\delta t, \D_B \delta t )\right]
+ \frac{\mathcal{J} (\delta t, \Phi^i)}{\A} \,\D_A 
\widetilde{W}_\mu{}^A (\Phi^i, \wD_B \Phi^i),
\label{G4}
\eeq
\end{widetext}
where $\widetilde{T}_\mu^A$, $\widetilde{W}_\mu^A$ and
$\mathcal{J}$ are given by~(\ref{FDs3}),~(\ref{WmuA}) 
and~(\ref{eq60}), respectively. Similarly, these arbitrary 
deformations are in agreement with those obtained in \cite{Rojas2022}
concerning the infinitesimal canonical transformations
induced by the first-class constraints under the Dirac-Bergmann
framework.

It is worth noting that, when we specialize these cumbersome 
transformations for a FRW geometry defined on the worldvolume, 
we get an astonishing reduction. For instance, in such a case,
$N^A = 0, \partial_A \Xt^\mu = 0$ and $\delta t^A = \delta \tau^A
= 0$ since there are no spatial deformations, so that, we recover 
the usual transformations early reported in~\cite{Rojas2020}. 

\section{Discussion}
\label{sec5}

In this paper we have developed a Hamilton-Jacobi framework 
for geodetic brane gravity described by the Regge-Teitelboim 
geometric model, in arbitrary co-dimension. After identifying 
all the HJPDE, this HJ scheme has been used as an alternative 
to analyze the phase space constraints content by identifying 
all the constraints. The subset of $2(p+1)$ constraints 
given by the densities (\ref{hami1a}), (\ref{hami1b}),
(\ref{hami2a}) and (\ref{hami2b}) furnish us with the involutive
Hamiltonians under the GB~(\ref{GB}) making the theory integrable
in a subspace of the phase space. We deduced similarly the 
subset of $2(N - p -1)$ non involutive constraints given by the 
densities (\ref{hami1c}) and (\ref{hami2c}) which help us to 
build a generalized bracket as well as to introduce a fundamental 
differential that governs the evolution in phase space. 
Unlike the Ostrogradsky-Dirac algorithm, in the HJ formalism 
the obtaining of all the constraints is based in the Frobenius 
integrability condition. Since the RT model is a theory affine 
in the accelerations of the extended object, we have couched 
our discussion based in an Ostrogradski-Hamilton approach since 
we are interested in maintaining the second-order geometric nature 
of this field theory rather than avoiding a surface term as in 
the case of general relativity. 
Of course, we could have chosen to use the auxiliary variables 
method to work with a first-order HJ approach \cite{pimentel2008} 
but, the price to pay is the enlargement of the phase space as well 
as the increase in the number of Hamiltonians that, in some sense, 
hide the original geometric properties of the theory. From the PDE 
viewpoint the analysis carried out shows that the theory is integrable 
in a subspace where the time parameter $t$ and the parameters $t^1, 
t^A$ and $\tau^A$, are the independent parameters related to the 
involutive Hamiltonians and where the physical evolution takes place. 
Guided by~\cite{Rojas2021} where it has been developed the HJ analysis 
for affine in acceleration theories, the present HJ study for the 
RT gravity has been performed. The complete characteristic 
equations, provided by the fundamental differential~(\ref{fdiff-2}), 
contain a complete set of arbitrary parameters where the 
equations of motion are included when the parameter $dt$ is 
solely considered. The remaining transformations associated 
with the other parameters represent gauge infinitesimal 
changes, see~(\ref{G1}-\ref{G4}). Indeed, the independence of 
the parameters $t$, $t^1$, $t^A$ and $\tau^A$, has been used to 
build the generator of gauge transformations given by 
$\mathbb{G}^{\text{\tiny can}}$. 
A particular issue of interest is the specialization of our findings 
to the cosmological scenario. Certainly, under a FRW geometry 
imposed on the worldvolume, and for the case of co-dimension 
one,~($i,j=1$), several functions of our development are 
simplified as $N = \sqrt{\dot{t}^2 - \dot{a}^2}$, $N^A = 0$
and $\B^A{}_i = 0$, to mention a few, so that several 
cumbersome geometrical structures like the Hamiltonians, test fields
and PB, are greatly reduced, which is also verified with the results
obtained in the Ostrogradsky-Hamilton framework~(see~\cite{Rojas2009,Rojas2020}, 
for details in the notation.).
However, with a view of the quantization process through either 
a path integral or a WKB scheme, it remains to explore the 
integration of the characteristic equation that obeys the 
generating function $S$, Eq.~(\ref{S}). This is because, as it is a 
singular theory, the associated extended action as well as the 
full set of HJPDE  where $S$ appears, should be carefully solved
and analyzed. Regarding this, some progress has been made for 
higher-order Lagrangians using the mentiones techniques and based
on this HJ approximation,~\cite{hasan2004,Muslih2002,hasan2014}. 
Work in this last direction, applied to this type of gravity, 
is in progress.

\section*{Acknowledgements}

ER dedicate this work to the memory of Prof. 
Riccardo Capovilla, a mentor, a colleague, and 
a dear friend. AAS acknowledges support from a 
CONACYT-M\'exico doctoral fellowship. ER and CC 
acknowledge encouragement from ProDeP-M\'exico, 
CA-UV-320: \'Algebra, Geometr\'\i a y 
Gravitaci\'on. ER is grateful to Alberto Molgado, 
Julio M\'endez-Zavaleta and Julio C. Natividad for 
discussions and helpful suggestions. Also, ER thanks 
partial support from Sistema Nacional de Investigadores, 
M\'exico.

\appendix

\section{Brane geometry in higher co-dimension}
\label{app1}

\subsection{$m$ embedded in $\mathcal{M}$}

Consider a $(p+1)$-dimensional manifold $m$ that represents 
the worldvolume, $m$, of a spacelike brane $\Sigma$. $m$ is 
embedded in a $N$-dim flat Minkowski spacetime, $\mathcal{M}$, 
with metric $\eta_{\mu\nu}
=\text{diag}(-1,1,...,1)$ ($\mu,\nu = 0,1,2, 
\ldots,N-1$). The worldvolume is described by the embedding 
functions $X^{\mu}(x^a)$ where $x^a$ are  local coordinates 
for $m$ ($a,b = 0,1,2, \ldots p$). The tangent vectors to 
$m$ are given by $e^\mu{}_a: = \partial X^\mu/\partial x^a$
which allow us to define an induced metric on $m$ as 
$g_{ab}:=\eta_{\mu\nu} e^{\mu}{}_a e^{\nu}{}_b=e_a 
\cdot e_b$. 
As usual, $g^{ab}$ denotes the inverse of $g_{ab}$, 
and $g = \det (g_{ab})$. Additionally, $m$ is assumed to be 
time-like so $g < 0$. The $i$th normal vector to $m$, 
$n^{\mu\,i}$, ($i,j=1,2,...,N-p-1$), is defined by the relations 
$n^i\cdot e_a=0$ and 
$n_i \cdot n_j=\delta_{ij}$. These expressions define the normal 
vectors up to a sign and a local $O(N-p-1)$ rotation. This 
gauge freedom requires a gauge field, known as \textit{twist 
potential}, given by $\omega_a{}^{ij} = - n^i \cdot 
\partial_a n^j$,~\cite{capovilla1995}. In the case of a 
hypersurface embedding, $i=1$ and the extrinsic twist vanishes 
identically. Further, $\nabla_a$ denotes the (torsionless) covariant
derivative compatible with $g_{ab}$. In this approach, 
the worldvolume gauge covariant derivative is given
by $\wt_a = \nabla_a - \omega_a$. 

The extrinsic curvature of $m$ is  $K_{ab}{}^i = 
- n^i \cdot \nabla_a \nabla_b X$, and the mean extrinsic 
curvature as its trace, $K^i = g^{ab} K_{ab}{}^i$.  
The corresponding Gauss-Weingarten (GW) equations are
\be 
\begin{aligned}
D_a e^\mu{}_b &= \Gamma^c_{ab} e^\mu{}_c - K_{ab}^i n^\mu{}_i,
\\
D_a n^{\mu\,i} & = K_{ab}^i g^{bc}e^\mu{}_c - \omega_a{}^{ij}n^\mu{}_j,
\end{aligned}
\ee
where $D_a = e^\mu{}_a D_\mu$ with $D_\mu$ is the covariant derivative
compatible with a possible background metric, say $\mathcal{G}_{\mu\nu}$, and
$\Gamma^a_{bc}$ are the connection coefficients compatible
with $g_{ab}$. The intrinsic and extrinsic geometries of the worldvolume 
$m$ are related by the integrability conditions. In this sense, the 
Gauss-Codazzi-Mainardi equations are
\beq 
0 &=& \R_{abcd} - K_{ac}{}^iK_{bd\,i} + K_{ad}{}^{i}K_{bc\,i},
\\
0 &=& \wt_a K_{bc}{}^i -\wt_b K_{ac}{}^i,
\\
0 &=& \Omega_{ab}{}^{ij} - K_{ac}{}^i K_b{}^{c\,j}
+ K_{bc}^i K_a{}^{c\,j}.
\eeq

\sk
In the same spirit, and depending of a particular viewpoint, we also
have integrability conditions describing the geometry of a brane 
at a fixed time, $\Sigma_t$, once we perform an ADM split.

\subsection{$\Sigma_t$ embedded in $m$}

If $\Sigma_t$ is embedded into $m$, $x^a = \chi^a (u^A)$, with $u^A$
being the local coordinates in $\Sigma_t$ and $A=1,2, \ldots, p$, the 
orthonormal basis is provided by $\{ \epsilon^a{}_A = \partial_a 
\chi^a, \eta^a \}$. This satisfies $g_{ab} \epsilon^a{}_A \eta^b 
= 0$, $g_{ab} \eta^a \eta^b = -1$ and $g_{ab} \epsilon^a{}_A 
\epsilon^b{}_B = h_{AB}$ where $h_{AB}$ is the spacelike metric 
associated with $\Sigma_t$. The corresponding GW equations are
\be 
\begin{aligned}
\nabla_A \epsilon^a{}_B &= \Gamma^C_{AB} \epsilon^a{}_C + k_{AB} \eta^a,
\\
\nabla_A \eta^a &= k_{AB} h^{BC} \epsilon^a{}_C,
\end{aligned}
\ee
where $\nabla_A = \epsilon^a{}_A \nabla_a$, $k_{AB} = k_{BA}$ is the 
extrinsic curvature of $\Sigma_t$ associated to the normal $\eta^a$
and $\Gamma^C_{AB}$ stands for the connection compatible with
$h_{AB}$. The intrinsic and extrinsic geometries for this embedding
should satisfy the integrability conditions
\beq 
\R_{abcd} \epsilon^a{}_A \epsilon^b{}_B \epsilon^c{}_C \epsilon^d{}_D
& = & R_{ABCD} - k_{AD} k_{BC} + k_{AC} k_{BD},
\\
\R_{abcd} \epsilon^a{}_A \epsilon^b{}_B \epsilon^c{}_C \eta^d
&=& \D_A k_{BC} - \D_B k_{AC},
\eeq
where $R_{ABCD}$ is the Riemann tensor associated with the spacelike 
manifold $\Sigma_t$ and $\D_A$ is the covariant derivative compatible 
with $h_{AB}$.

\subsection{$\Sigma_t$ embedded in $\mathcal{M}$}

If $\Sigma_t$ is embedded into $\mathcal{M}$, $x^\mu = X^\mu (u^A)$,
the orthonormal basis is provided by $\{ \epsilon^\mu{}_A = 
\partial_A X^\mu, \eta^\mu, n^\mu{}_i \}$. This satisfies 
$\epsilon_A \cdot \eta = \epsilon_A \cdot n_i = \eta \cdot n_i 
= 0$, $ \eta \cdot \eta = -1$, $n_i \cdot n_j = \delta_{ij}$ and 
$ \epsilon_A \cdot \epsilon_B = h_{AB}$. The corresponding GW 
equations are
\be 
\begin{aligned}
D_A \epsilon^\mu{}_B &= \Gamma^C_{AB} \epsilon^\mu{}_C + k_{AB} 
\eta^\mu - L_{AB}^i \,n^\mu{}_i,
\\
D_A \eta^\mu &= k_{AB} \epsilon^{\mu\,B} - L_A{}^i n^\mu{}_i,
\\
D_A n^\mu{}_i & = L_{AB}^i  \epsilon^{\mu\,B} - L_A{}^i \eta^\mu 
+ \sigma_A{}^{ij} n^\mu{}_j, 
\end{aligned}
\ee
where $D_A = \epsilon^\mu {}_A D_\mu$ and $D_\mu$ being the 
background covariant derivative, $L_{AB}^i = L_{BA}^i$ is the 
extrinsic curvature of $\Sigma_t$ associated with the normal 
$n^\mu{}_i$. Additionally, we have introduced $L_A{}^i := 
\epsilon^a{}_A \eta^b K_{ab}^i$ and $\sigma_A{}^{ij} := 
\epsilon^a{}_A \omega_a{}^{ij}$. Further, we can rewrite 
$L_{AB}^i$ and $L_A{}^i$ as $L_{AB}^i = - n^i \cdot \D_A \D_B X$ 
and $L_A{}^i = \eta \cdot \wD_A n^i$, respectively, where
$\wD_A$ is the covariant derivative acting on the normal indices
associated with the connection $\sigma_A{}^{ij}$, $\Dt_A \Phi^i = 
\D_A \Phi^i - \sigma_A{}^{ij}\Phi_j$.

The intrinsic and extrinsic geometries for this embedding 
should satisfy the integrability conditions
\be 
\begin{aligned}
0 &= - R_{ABCD} - k_{AC} k_{BD} + k_{BC} k_{AD} + L_{AC}^i L_{BD\,i}
\\
& - L_{BC}^i L_{AD\,i},
\\
0&= \D_A k_{BC} - \D_B k_{AC} + L_A{}^i L_{BC\,i} - L_B{}^i L_{AC\,i},
\\
0 &= \wD_A L_{BC}^i - \wD_B L_{AC}^i + L_A{}^i k_{BC} - L_B{}^i k_{AC},
\\
0 &= \wD_A L_B{}^i - \wD_B L_A{}^i + L_A{}^{C\,i}k_{BC} - L_B{}^{C\,i}
k_{AC},
\\
0 &= - \Omega_{AB}{}^{ij} + L_A{}^{C\,i}L_{BC}^j - L_B{}^{C\,i}L_{AC}^j
\\
& - L_A{}^i L_B{}^j + L_B{}^i L_A{}^j,
\end{aligned}
\label{int-cond}
\ee
where $\Omega_{AB}{}^{ij} := \D_B \sigma_A{}^{ij} - \D_A \sigma_B{}^{ij}
+ \sigma_A{}^{ik} \sigma_{B\,k}{}^j - \sigma_B{}^{ik}\sigma_{A\,k}{}^j$
is the curvature tensor associated with the gauge field $\sigma_A{}^{ij}$.
Additionally, by contracting some of integrability conditions \ref{int-cond},
we obtain the following relationships which are useful in the calculations
\beq 
\D_B (k^{AB} - h^{AB} k) &=& \left( h^{AB} L_i - L^{AB}_i \right)L_B{}^i,
\label{id20}
\\
\wD_B \left( L^{AB}_i - h^{AB} L_i \right) &=& \left( h^{AB}k - k^{AB}
\right)L_{B\,i}.
\label{id21}
\eeq

\sk
\section{Functional derivatives of the Hamiltonians}
\label{app2}

Here we present the functional derivatives of the 
Hamiltonian constraint functions  
\begin{widetext}
\be 
\begin{aligned}
\frac{\delta H^{'P}_\lambda}{\delta \dot{X}^\mu}
= \lambda\,P_\mu , \qquad \qquad \qquad \quad \quad
\qquad
& \qquad \qquad
\frac{\delta H^{'P}_\lambda}{\delta P_\mu}
= \lambda\,\dot{X}^\mu,
\\
\frac{\delta H^{'P}_{\vec{\lambda}}}{\delta X^\mu}
= - \partial_A (\lambda^A P_\mu)
= - \pounds_{\vec{\lambda}}P_\mu,
\qquad \,
& \qquad \qquad
\frac{\delta H^{'P}_{\vec{\lambda}}}{\delta P_\mu}
= \lambda^A \partial_A X^\mu
=  \pounds_{\vec{\lambda}}X^\mu,
\\
\frac{\delta H^{'P}_{\vec{\phi}}}{\delta \dot{X}^\mu}
= -  \phi^i \frac{\sqrt{h}}{N^2}L_i\,\eta_\mu
-  \phi^i \frac{\sqrt{h}}{N^2} k\,n_{\mu\,i},\,\,
& \qquad \qquad
\frac{\delta H^{'P}_{\vec{\phi}}}{\delta P_\mu}
= \phi^i\,n^\mu{}_i,
\\
\frac{\delta H^{'P}_{\vec{\phi}}}{\delta X^\mu}
= \D_A \widetilde{w}^A _\mu (\phi^i),\,\,\, \qquad \qquad \qquad \quad 
& \qquad \qquad 
\end{aligned}
\ee
where
\be 
\widetilde{w}^A _\mu  = - \phi^i \frac{1}{N} \B^A{}_i \eta_\mu 
- \phi^i \frac{\h}{N} (L^{AB}_i - h^{AB} L_i) \epsilon_{\mu\,B}
- \frac{\h}{N^2} (\phi^i N^A k + \phi^i \D^A N - N \wD^A \phi^i)
n_{\mu\,i},
\ee
with $\B^A{}_i$ given by (\ref{BAi}). Further, 
$\pounds_{\vec{\lambda}}$ stands for the Lie derivative 
with respect to the vector field $\vec{\lambda}$. Similarly,
\be
\begin{aligned} 
\frac{\delta H^{'p}_\Lambda}{\delta p_\mu}
&=
\Lambda\,\dot{X}^\mu
\qquad \qquad \qquad \qquad \qquad \qquad 
\frac{\delta H^{'p}_\Lambda}{\delta P_\mu}
=
2\Lambda N^A\D_A \dot{X}^\mu 
+ \Lambda (N^2 h^{AB} - N^A N^B) \D_A \D_B X^\mu,
\\
\frac{\delta H^{'p}_\Lambda}{\delta \Xt^\mu}
&=
\Lambda\,p_\mu 
+ 2\Lambda \,(P \cdot \D_A \dot{X})\, \epsilon_{\mu}{}^A 
- \D_A (2\Lambda N^A\,P_\mu) - 2\Lambda N h^{AB} (P \cdot 
\D_A \D_B X)\, \eta_\mu
- 2 \Lambda N^A (P \cdot \D_A \D_B X) \epsilon_\mu{}^B
\\
& - \Lambda \sh \left( \frac{1}{2}\Theta
+ \Psi \right) \eta_\mu
+  \Lambda \sh (G^{ABCD}L_{AB}^i k_{CD})\,n_{\mu\,i}
- 2 \frac{\h}{N} \left[ \wD_A (\Lambda N L^A{}_i) 
- \Lambda N L^{AB}_i k_{AB}
\right]n_\mu{}^i,
\\
\frac{\delta H^{'p}_\Lambda}{\delta X^\mu}
&= \D_A \widetilde{T}_\mu ^A (\Lambda, \D_B \Lambda),
\end{aligned}
\end{equation}
where
\beq
\widetilde{T}_\mu^A &:=&  2\Lambda N \B^A{}_j L^j
\,\eta_\mu 
- \Lambda \h \,N^A \left( \frac{1}{2} \Theta + \Psi \right)
\eta_\mu
- 2 \Lambda \h \left( N^A L^B{}_i L^i + N L^A{}_{C\,i}
L^{BC\,i} \right)\epsilon_{\mu\,B}
\n
\\
&-& \Lambda \h N \left( \frac{1}{2} \Theta + \Psi \right)
\epsilon_\mu{}^A 
+ 2\Lambda \h N \left( \wD^A n^i \cdot 
\wD_B n_i \right) \epsilon_{\mu}{}^B
+  \Lambda \h N^A (G^{BCDE} L_{BC}^i k_{DE})\,
n_{\mu\,i}
\n
\\
&-& 2 \h \left[ \wD_B (\Lambda N L^{AB}_i) - 
\Lambda N k^{AB} L_{B\,i} \right]
n_{\mu}{}^i 
- 2\h \frac{N^A}{N}
\left[ \wD_B (\Lambda N L^{B}{}_i) - 
\Lambda N k^{BC} L_{BC\,i} \right]
n_\mu{}^i
\n
\\
&+& \D_B \left[ \Lambda (N^2 h^{AB} - N^A N^B)
P_\mu \right]
+ \h \,\D_B \left[\Lambda N \,(G^{ABCD} \Pi_{\mu\nu}
\D_C \D_D X^\nu)\right],
\label{FDs3} 
\eeq
with $\Theta$ and $\Psi$ given by~(\ref{Theta}) and~(\ref{Psi}),
respectively. Additionally,
\beq 
\frac{\delta H^{'p}_{\vec{\Lambda}}}{\delta \dot{X}^\mu}
&=& - \partial_A (\Lambda^A P_\mu) = - 
\pounds_{\vec{\Lambda}}P_\mu,
\qquad \qquad
\frac{\delta H^{'p}_{\vec{\Lambda}}}{\delta P_\mu} 
= \Lambda^A \partial_A \dot{X}^\mu = \pounds_{\vec{\Lambda}}
\dot{X}^\mu,
\\
\frac{\delta H^{'p}_{\vec{\Lambda}}}{\delta X^\mu}
&=& - \partial_A (\Lambda^A p_\mu) = - 
\pounds_{\vec{\Lambda}}p_\mu,
\qquad \qquad
\frac{\delta H^{'p}_{\vec{\Lambda}}}{\delta p_\mu} 
= \Lambda^A \partial_A X^\mu = \pounds_{\vec{\Lambda}}X^\mu,
\eeq
and 
\be
\begin{aligned}
\frac{\delta H^{'p}_{\vec{\Phi}}}{\delta p_\mu} 
& = \Phi^i n^\mu{}_i,
\qquad \qquad \qquad \qquad \qquad \qquad \qquad
\frac{\delta H^{'p}_{\vec{\Phi}}}{\delta P_\mu} = N^A \mathcal{D}_A
(\Phi^i n^\mu{}_i) = \mathcal{L}_{\vec{N}} ( \Phi^i n^\mu{}_i),
\\
\frac{\delta H^{'p}_{\vec{\Phi}}}{\delta \dot{X}^\mu} 
& = \frac{\h}{N}L_i (\wD_A \Phi^i)\,\epsilon_\mu{}^A 
+ \frac{\h}{N} (\wD_A \wD^A \Phi^i)\,n_{\mu\,i} 
+ \frac{\h}{N} \Phi^i L^A{}_i L_A{}^j \, n_{\mu\,j}
+ \frac{\h}{2N} \Phi^i 
\Theta\,n_{\mu\,i}
\\
\frac{\delta H^{'p}_{\vec{\Phi}}}{\delta X^\mu} &= 
\D_A \widetilde{W}^A_\mu (\Phi^i, \wD_B \Phi^i, \wD_B \wD_C \Phi^i),
\end{aligned}
\ee
where 
\beq
\widetilde{W}^A_\mu &=&  
\h
\left[ \left( L^{AB}_i - h^{AB}L_i \right)( \wD_B \Phi^i) \,\eta_\mu  
+ \frac{1}{\sh} \B^A{}_i (\wD_B \Phi^i) \,\epsilon_\mu{}^B
+ L^B{}_i (\wD^A \Phi^i)\,\epsilon_{\mu\,B}
- L^B{}_i (\wD_B \Phi^i)\,\epsilon_\mu{}^A
\right.
\n
\\
&+& \left. \Phi^i \frac{1}{N} (N^A L_{B\,i} + N L^A{}_{B\,i})L^B{}_j
\,n_\mu{}^j - k^{AB} (\wD_B \Phi^i )\,n_{\mu\,i}
+ \Phi^i \D_B \left( k^{AB} - h^{AB}k\right)\,n_{\mu\,i}
+ \Phi^i \frac{N^A}{2N} \Theta \, n_{\mu\,i} 
\right. 
\n
\\
&+& \left. \frac{N^A}{N} (\wD_B \wD^B \Phi^i) \,n_{\mu\,i} \right].
\label{WmuA}
\eeq
\end{widetext}

\begin{widetext}

\section{Extended Poisson brackets among the Hamiltonians}
\label{AppC}

We start with
\beq 
\{ H^{'P}_\lambda , H^{'P}_{\lambda'} \}
&=& 
0,
\\
\{ H^{'P}_\lambda , H^{'P}_{\vec{\lambda}} \}
&=& 
H^{'P}_{\vec{\lambda_1}} \qquad \qquad \qquad \lambda_1^A 
= \lambda \lambda^A,
\\
\{ H^{'P}_\lambda , H^{'P}_{\vec{\phi}} \} &=& 
H^{'P}_{\vec{\phi_1}} \qquad \qquad\qquad \,\phi^i_1 = 
\lambda \phi^i ,
\\
\{ H^{'P}_{\vec{\lambda}}, H^{'P}_{\vec{\lambda'}} \}
&=& 
0,
\\
\{ H^{'P}_{\vec{\lambda}}, H^{'P}_{\vec{\phi}} \}
&=& 
0,
\\
\{ H^{'P}_{\vec{\phi}}, H^{'P}_{\vec{\phi \,'}} \}
&=& 0.
\eeq

Similarly,
\beq 
\{ H^{'P}_\lambda, H^{'p}_\Lambda \}
&=& - H^{'P}_{\lambda_1} - H^{'p}_{\Lambda_1} 
\qquad \qquad \qquad \qquad 
\lambda_1 := 2\Lambda N^A \D_A \lambda,
\qquad \,\,\,\Lambda_1 := \lambda \Lambda,
\\
\{ H^{'P}_\lambda, H^{'p}_{\vec{\Lambda}} \}
&=& - H^{'P}_{\lambda_2}
\qquad \qquad \qquad \qquad \qquad \quad 
\lambda_2 := \Lambda^A \D_A \lambda,
\\
\{ H^{'P}_\lambda , H^{'p}_{\vec{\Phi}} \}
&=&
H^{'P}_{\vec{\phi_2}} 
\qquad \qquad \qquad\qquad \qquad \quad\,\,\,\,
\phi^i_2 := \lambda N^A \wD_A \Phi^i,
\\
\{ H^{'P}_{\vec{\lambda}}, H^{'p}_\Lambda \}
&=&  H^{'P}_{\lambda_3} - H^{'P}_{\vec{\lambda_2}} 
- H^{'p}_{\vec{\Lambda_2}} 
\qquad \qquad \quad\,\,\,\, 
\lambda_3 := \lambda^A \D_A \Lambda,
\qquad \lambda_2^A := 2\Lambda N^B \D_B \lambda^A,
\qquad \Lambda_2^A := \Lambda \lambda^A,
\\
\{ H^{'P}_{\vec{\lambda}}, H^{'p}_{\vec{\Lambda}} \}
&=& H^{'P}_{[\vec{\lambda},\vec{\Lambda}]},
\\
\{ H^{'P}_{\vec{\lambda}} , H^{'p}_{\vec{\Phi}} \}
&=&
H^{'P}_{\vec{\phi_3}} 
\qquad \qquad \qquad\qquad\qquad \quad \,\,\,\,
\phi^i_3 := \lambda^A \wD_A \Phi^i,
\\
\{ H^{'P}_{\vec{\phi}}, H^{'p}_\Lambda \}
&=& 
H^{'P}_{\lambda_4} + H^{'P}_{\vec{\lambda_3}}
- H^{'P}_{\vec{\phi_4}} - H^{'p}_{\vec{\Phi_1}},
\\
\{ H^{'P}_{\vec{\phi}}, H^{'p}_{\vec{\Lambda}} \}
&=& H^{'P}_{\lambda_5} - H^{'P}_{\vec{\lambda_4}}
- H^{'P}_{\vec{\phi_5}}
\qquad \qquad \quad\,\,
\lambda_5 := \frac{1}{N}\Lambda^A L_A{}^i\,\phi_i, 
\qquad
\lambda_4^A := \frac{1}{N} \Lambda^B (N^A L_B{}^i 
+ N L_B{}^{A\,i})\,\phi_i,
\\
\{ H^{'P}_{\vec{\phi}} , H^{'p}_{\vec{\Phi}} \}
&=& \A,
\eeq
where
\beq 
\lambda_4 &:=& \Lambda \left( \frac{N^A}{N} L_A{}^i - 
L^i \right)\phi_i,
\\
\lambda^A_3 &:=& \Lambda \left( N^A L^i - N^B L_B{}^{A\,i} 
- \frac{N^A N^B}{N} L_B{}^i\right)\phi_i,
\\
\phi_4^i &:=& \Lambda Nk \,\phi^i  - N^A \D_A \Lambda
\,\phi^i
+ \Lambda N^A \wD_A \phi^i,
\\
\Phi_1^i &:=& \Lambda \phi^i,
\\
\phi_5^i &:=& \Lambda^A \wD_A \phi^i,
\eeq
and $\A$,~Eq.~(\ref{eq61}), is given in terms of~(\ref{Aij}).

\bk
Finally, we have 

\beq 
\{ H^{'p}_\Lambda, H^{'p}_{\Lambda'} \} &=& 
H^{'P}_{\lambda_6} + H^{'P}_{\vec{\phi_6}}
\qquad \qquad \qquad \qquad \qquad \quad
\lambda_6 := (N^2 h^{AB} - N^A N^B)(\Lambda \D_A \D_B \Lambda'
- \Lambda' \D_A \D_B \Lambda),
\\
\{ H^{'p}_\Lambda , H^{'p}_{\vec{\Lambda}} \}
&=& - H^{'P}_{\lambda_7} + H^{'P}_{\vec{\lambda_6}} 
\qquad \qquad \quad \qquad \qquad \quad
\lambda^A_6 := \Lambda (N^2 h^{BC} - N^B N^C)(\D_B \D_C \Lambda^A
+ R_{BDC}{}^A \Lambda^D), 
\\
\{ H^{'p}_{\Lambda} , H^{'p}_{\vec{\Phi}} \}
&=&  H^{'P}_{\lambda_8} -  H^{'P}_{\vec{\lambda_7}}
- H^{'P}_{\vec{\phi_7}} - H^{'p}_{\Lambda_2}
+ H^{'p}_{\vec{\Lambda_3}} 
\n
\\
&+& H^{'p}_{\vec{\Phi_2}} + \mathcal{J} (\Lambda, \Phi^i),
\\ 
\{ H^{'p}_{\vec{\lambda}}, H^{'p}_{\vec{\Lambda'}}\} &=&
H^{'p}_{[\vec{\Lambda}, \vec{\Lambda\,'}]}, 
\\
\{ H^{'p}_{\vec{\Lambda}} , H^{'p}_{\vec{\Phi}} \}
&=&
- H^{'P}_{\lambda_9} + H^{'P}_{\vec{\lambda_8}}
+ H^{'P}_{\vec{\phi_8}} - H^{'p}_{\Lambda_3}
+ H^{'p}_{\vec{\Lambda_4}}
\n
\\
&+&  H^{'p}_{\vec{\Phi_3}},
\\
\{ H^{'p}_{\vec{\Phi}} , H^{'p}_{\vec{\Phi\,'}} \}
&=& \F,
\eeq
where $\F$ and $\mathcal{J}$,~Eqs.~(\ref{eq62}) and~(\ref{eq60}), 
are given in terms of~(\ref{FAij}) and~(\ref{J}), and
\beq 
\lambda_7 &:=& \Lambda^A \D_A \Lambda,
\\
\phi_6^i &:=&
2N^3 h^{AB} L_A{}^i (\Lambda \D_B \Lambda' - 
\Lambda' \D_B \Lambda),
\\
\lambda_8 &:=& \frac{\Lambda}{N} \left[ 
(N^2 h^{AB} + N^A N^B) \left( k_{AC} L_B{}^{C\,i}\, \Phi_i -
L_A{}^i \wD_B \Phi_i \right) - 2N^A \D_A (N^B L_B{}^i\, \Phi_i)
\right],
\\
\lambda_7^C &:=& \Lambda \left\lbrace
\frac{1}{N}(N^2 h^{AB} + N^A N^B) (N L_A{}^i k_B{}^C 
+ N^C L_{AD}^i k_B{}^D) \Phi_i
\right.
\n
\\
&+& \left. \frac{1}{N}(N^2 h^{AB} - N^A N^B) (N L^C{}_{B\,i} 
+ N^C L_{B\,i}) \wD_A \Phi_i - \frac{2}{N} \left[ N N^A 
\D_A (N^B L_B{}^{C\,i}\Phi_i) + N^C N^A \D_A (N^B L_B{}^i 
\Phi_i)\right]
\right.
\n
\\
&-& \left. 2 \frac{N^A N^B}{N} (N L_B{}^{C\,i} + N^C L_B{}^i )\,
\wD_A \Phi_i \right\rbrace,
\\
\phi_7^i &:=& \Lambda \left[ (N^2 h^{AB} + N^A N^B)\left( 
\wD_A n^i \cdot \wD_B n^j\right)\Phi_j - (N^2 h^{AB} - N^A N^B) \wD_A
\wD_B \Phi^i \right],
\\
\Lambda_2 &:=& \frac{\Lambda}{N} N^A L_A{}^i\,\Phi_i,
\\
\Lambda_3^A &:=& \frac{\Lambda}{N} N^B (N L_B{}^{A\,i} +
N^A L_B{}^i)\,\Phi_i,
\\
\Phi_2^i &:=& \Lambda N^A \wD_A \Phi^i,
\eeq
and
\beq
\lambda_9 &:=& \frac{1}{N} \Lambda^A \D_A (N^B L_B{}^i \Phi_i)
+ \frac{1}{N}\Lambda^A N^B(L_A{}^i \wD_B \Phi_i - k_{AC}
L_B{}^{C\,i}\,\Phi_i),
\\
\lambda_8^C &:=& \frac{1}{N}\Lambda^A \D_A (N^B L_B{}^{C\,i} 
\Phi_i) + \frac{1}{N}\Lambda^A N^C \D_A (N^BL_B{}^i \Phi_i)
+ \frac{1}{N} \Lambda^A N^B (N^C L_A{}^i + N L_A{}^{C\,i}) 
\wD_B \Phi_i
\n
\\
&-& \frac{1}{N} \Lambda^A N^B(N k_A{}^C L_B{}^i + N^C k_{AD} 
L_B{}^{D\,i}) \Phi_i,
\\
\phi_8^i &:=& \Lambda^A \wD_A (N^B \wD_B \Phi^i)
- \Lambda^A N^B ( \wD_A n^i \cdot \wD_B n^j) \Phi_j,
\\
\Lambda_3 &:=& \frac{1}{N} \Lambda^A L_A{}^i\Phi_i,
\\
\Lambda_4^B &:=& \frac{1}{N} \Lambda^A (N^B L_A{}^i
+ N L_A{}^{B\,i}) \Phi_i,
\\
\Phi_3^i &:=& \Lambda^A \wD_A \Phi^i. 
\eeq
\end{widetext}

\subsection{Structures $\widetilde{A}_{ij}$, $\widetilde{\F}^A_{ij}$, 
$\widetilde{\F}_i$ and $\widetilde{\mathcal{J}}$} 
\label{subAppb}

We provide here the explicit form of the structures appearing 
in~(\ref{eq60})~(\ref{eq61}) and~(\ref{eq62}),
\begin{widetext}
\beq 
\widetilde{\A}_{ij} &:=& \frac{\h}{N} \left[ \left( L^{AB}_i 
L_{AB\,j} - L_i L_j \right) - \frac{1}{2} \delta_{ij}\,\Theta
\right],
\label{Aij}
\\
\widetilde{\F}^A_{ij} 
&:=& 
\h ( L^{AB}_l - h^{AB} L_l ) (\delta_{ij}L_B{}^l + \delta_i{}^l 
L_{B\,j} + \delta_j{}^l L_{B\,i}),
\label{FAij}
\\
\widetilde{\F}_i(\Phi^j, \wD_A \Phi^j)
&:=& 
\h \left\lbrace  2 \left( L^{AB\,k} - h^{AB}L^k\right) L_{A\,kij} \wD_B \Phi^j 
- \wD_A \left[ \left( L^{AB\,k} - h^{AB}L^k\right) 
L_{A\,kij} \right] \Phi^j \right\rbrace,
\\
\widetilde{\mathcal{J}} (\Phi^i, \wD_A \Phi^i) &:=& \h N \left[ 
L_i \left( \Theta + \Psi \right) - 2L_{A\,i} \D_B \left(k^{AB} 
- h^{AB} k \right) 
-  L^{AB}_i  \left( \wD_A n^j
\cdot \wD_B n_j \right)
\right] \Phi^i
\n
\\
&+& \h N^A\,G_A{}^{BCD}L_B{}^i \wD_C \wD_D \Phi_i.
\label{J}
\eeq
Here, we have introduced $L_{A\,kij} := L_{A\,k} \delta_{ij} 
+ L_{A\,i}\delta_{jk} + L_{A\,j}\delta_{ki}$. Notice further
that $\widetilde{\A}_{ij} = \widetilde{\A}_{ji}$ and 
$\widetilde{\F}^A_{ij} = \widetilde{\F}^A_{ji}$.
\end{widetext}



\end{document}